\documentclass[aps,pre,twocolumn,showpacs,amssymb]{revtex4}
\usepackage[dvips]{graphicx}
\usepackage{dcolumn}
\usepackage{bm}

\def\gs{\gtrsim}
\def\ls{\lesssim}
\def\be{\begin{equation}}
\def\en{\end{equation}}                  
\def\p{\partial} 

\def\bea{\begin{eqnarray}}
\def\ena{\end{eqnarray}}
\newcommand{\ppp}[3]{{\bigg(}\frac{\partial {#1}}{\partial {#2}}{\bigg )}_{#3}}
\def\ge{> \kern -12pt \lower 5pt \hbox{$\displaystyle =$}}
\def\le{< \kern -12pt \lower 5pt \hbox{$\displaystyle =$}}
\def\gs{> \kern -12pt \lower 5pt \hbox{$\displaystyle{\sim}$}}
\def\ls{< \kern -12pt \lower 5pt \hbox{$\displaystyle{\sim}$}}

\renewcommand{\theequation}{\arabic{section}.\arabic{equation}}

\begin{document}
\title{Thermoacoustic  effects in supercritical fluids near the 
critical point: Resonance, piston effect, 
and acoustic emission and reflection}
\author{Akira Onuki}
\affiliation{Department of Physics, Kyoto University, Kyoto 606-8502}

\begin{abstract}
We present a general theory of  thermoacoustic 
phenomena in supercritical fluids 
near the critical point in a one-dimensional cell. 
We take into account the effects of 
the heat conduction in the boundary walls  
and the bulk viscosity near the critical point.
We  introduce a coefficient $Z(\omega)$ characterizing 
 reflection of sound with frequency $\omega$ at the boundary. 
As applications, we 
examine  the  acoustic  eigenmodes in the cell,  
 the response  to time-dependent perturbations, 
sound emission and  reflection 
at the boundary. Resonance and rapid 
adiabatic changes are noteworthy. 
In these processes, the role of 
the thermal diffusion layers is enhanced near the critical point 
because of the strong critical divergence of 
the  thermal expansion. 
\end{abstract}
\maketitle

\section{Introduction}

In highly compressible fluids, 
adiabatic changes take place  with propagation of 
sounds and are much faster than the  thermal 
diffusion  \cite{Onukibook}.  When  a one-component 
fluid is heated or cooled at a boundary, 
a thermal diffusion layer  
expands or shrinks to emit  sounds, which then  
cause adiabatic changes in the interior 
(the thermal  piston effect). 
The density change in the boundary layer 
 is enhanced near the gas-liquid critical point 
because of the strong critical growth of the isobaric 
thermal expansion. 
As a result, thermal equilibration times become shorter 
near the critical point  
at fixed volume, despite the fact that the thermal 
diffusion constant $D$ tends to zero  at the criticality   
\cite{Straub,Ferrell,Gammon,Beysens,Be,Klein,Wilkinson,Straubp,Zhong,Kogan}.  
If the boundary temperature 
is slightly changed, temperature homogenization 
occurs throughout the cell  
on  a scale of the piston time \cite{Ferrell}, 
\be 
t_1= L^2/4(\gamma-1)^2D, 
\en 
where $L$ is the 
cell length, $D$ is the thermal diffusion constant, 
  and $\gamma=C_p/C_V$ 
is the specific-heat ratio growing near the critical point. 
See Appendix A for a derivation  of Eq.(1.1). 
The time $t_1$ is  much shorter  than 
the isobaric equilibration time $L^2/4D$ by the  factor 
  $(\gamma-1)^{-2} \ll 1$.   
For example, we have $t_1= 6.3\times 10^4\epsilon^{1.65}$sec 
for CO$_2$ with $L=1$cm. 
Hereafter, $\epsilon=T/T_c-1$ is the reduced temperature 
near the critical point. 
The early  experiments   detected  
 slow  temperature and density changes in the interior 
on time scales  much longer than the acoustic time 
$L/c$.  The fast acoustic  processes   
were  examined  by numerical simulations of 
the   hydrodynamic equations 
of compressible fluids  
\cite{Beysens,ZA,Maekawa}.

 Ferrell  and Hao \cite{Hao} found  relevance of 
 the heat conduction in the boundary 
walls in transient heat transport.  That is, 
the thermal boundary condition of a cell containing 
a near-critical fluid 
crosses over from the isothermal to insulating one 
even for a metal boundary wall 
due to the critical  divergence 
of the effusivity of the fluid 
\cite{effusive}.  The formula (1.1) should then be 
modified, because it is based on the isothermal 
boundary condition.  More recently,  
Carl$\grave{\rm e}$s and Dadzie \cite{Carles0,Carles} 
found that the bulk viscosity, which grows  strongly near 
the critical point, can affect 
the hydrodynamics in the thermal diffusion layer.  
Gillis {\it et al.} \cite{Moldover} 
performed experiments of acoustic resonance in xenon,  
where the frequency and attenuation of the 
resonating  modes  were measured. 
For such long wavelength sounds, 
the heat conduction 
at the boundary is  the dominant 
damping  mechanism 
relatively far from the critical point, 
while  the viscous effect in the bulk becomes more 
important closer to the critical point. 
They also presented thorough theoretical 
analysis of their  data. 
The critical growth of 
the effusivity   and the  bulk viscosity  
of the fluid  both serve to 
suppress the  boundary damping, as confirmed 
experimentally and theoretically. 
Very recently, Miura {\it et al.} \cite{Ohnishi,Miura} measured 
acoustic  density changes with precision 
of order $10^{-7}$g$/$cm$^3$ in  
near-critical CO$_2$  on  the acoustic time scale  
using a ultra-sensitive interferometer. 
They detected emission and traversal  of 
sound pulses  with width 
of order $10\mu$sec, which were   broadened 
as they moved through  the cell and interacted  
with the boundary walls. Some of their data agreed with 
predictions, but most data remain unexplained. 
Afterwords, part of the measured time-evolution of the density 
was numerically reproduced 
by Carl$\grave{\rm e}$s,  neglecting  the bulk viscosity 
 \cite{Carlesnew}.

Some unique aspects 
of  the  supercritical hydrodynamics
have been revealed by  experiments 
\cite{Meyer,Azuma,jet} 
and by simulations 
\cite{Maekawa,Sakir0,Chiwata,Soboleva,Sakir,Accary}. 
In  Rayleigh-B$\acute{\rm e}$nard 
convection,  
 overall temperature changes are  induced by 
plume arrivals at the 
boundary walls  due to the piston effect, 
leading to  overshoot behavior 
observed  in experiments of $^3$He 
near its critical point \cite{Meyer,Chiwata,Sakir}.
Significant noises of the adiabatic temperature changes 
 were predicted  in  
turbulent convective states \cite{Chiwata}, 
though not yet measured systematically. 
Recently three-dimensional simulations were performed 
\cite{Accary}.  Due to large thermal expansion 
in supercritical fluids, jet-like fluid flow 
has been observed around a heated boundary \cite{Azuma,jet}. 
In these processes, the  plume motions 
governed by the shear viscosity are 
strongly influenced by large thermal expansion 
around a heater and by rapid adiabatic 
density and temperature changes achieved by sound 
propagation.

In this paper, we aim to give detailed 
analysis of the linear hydrodynamics of supercritical fluids 
near the critical point in a one-dimensional cell. 
We take into account the effects of the decreasing 
 effusivity ratio \cite{effusive} 
and the growing bulk viscosity. 
In Section II, we will decompose fluid motions 
into sound modes and thermal diffusion modes with frequency $\omega$. 
These two modes are mixed at the boundary 
under given  boundary conditions, 
leading to various thermoacoustic 
phenomena. In Section III, 
 we will study the acoustic eigenmodes 
determined 
to confirm the calculations by Gillis {\it et al.}\cite{Moldover} 
in the simpler one-dimensional geometry.  
We will also examine the response of the fluid to 
various time-dependent perturbations.  
Resonance is induced 
 when the  frequency of the perturbation 
is close to one of the eigenfrequencies, while 
nearly uniform adiabatic changes are caused 
in the interior  due to the piston effect 
at much lower  frequencies. 
 We will also examine 
sound emission and reflection  at the boundary.
In Appendix A, we will present a simple theory 
of the piston effect, which can  be a  
starting  point to understand the complicated 
calculations  in the text. In  Appendix B, the critical 
behavior of one-component fluids used in the text 
will be  summarized.

\setcounter{equation}{0}
\section{Theoretical Background}

\subsection{Linear hydrodynamics}

Near the critical point, we treat 
the hydrodynamic deviations with spatial scales 
much exceeding  the thermal correlation length $\xi$, 
but the typical frequency $\omega$ can be higher  than the 
relaxation rate  of the critical fluctuations 
$t_\xi^{-1} \propto \epsilon^{1.89}$ \cite{Swinney}. 
  For such high frequencies $\omega$, 
 the bulk viscosity $\zeta$ behaves as 
 $1/\omega$, 
while   $\zeta\propto \epsilon^{-1.67}$ for $\omega t_\xi<1$ 
\cite{FB1,Onuki97,FolkMoser}. 
See Appendix B for more details. 
The other transport coefficients may be treated to be 
independent of $\omega$ \cite{Onukibook}. 
The critical singularity of 
the shear viscosity $\eta$ is negligible small, while  
the critical growth of the thermal conductivity $\lambda$ 
arises from the convective motions of the critical fluctuations 
taking place on a short time scale of order 
$\rho\xi^2/\eta$. Here we assume $ \omega\ll \eta/\rho\xi^2$.

The  mass density, the temperature, 
the entropy (per unit mass), and the pressure are written as 
$\rho$,  $ T$,  $ s$, and $ p$,  
respectively, with  their  small deviations  being 
$\delta\rho$,  $\delta T$,  $\delta s$, and $\delta p$. 
The velocity in  the $x$ direction 
is written as $v$.  These deviations  depend on time $t$ as 
$\exp(i\omega t)$  and vary in space along the 
$x$ axis.  We may assume $\omega>0$ without loss of generality 
(see Eq.(3.1)). 
These deviations  may be regarded  as the Fourier transformations 
 of the space-time dependent deviations 
with respect to time ($=\int dt e^{-i\omega t}
(\cdots)$). They obey the linear equations \cite{Landau},  
\bea 
i\omega \delta\rho &=& -\rho v',\\
i\omega \rho v &=& -\delta p' + \rho\nu_{\ell}  v'',\\
i\omega \rho T\delta s &=& \lambda \delta T''.
\ena
Here  the prime denotes the differentiation 
with respect to $x$. We have two dissipative coefficients;
one is the thermal conductivity  $\lambda$ and the other  is    
\be 
\nu_{\ell}=(\zeta+4\eta/3)/\rho, 
\en 
where  $\zeta$ and $\eta$ are  
the bulk and shear viscosities, respectively. 
Using the thermodynamic derivatives we may 
 express $\delta s$ and $\delta p$ in terms of 
 $\delta \rho$ and $\delta T$ as 
\bea 
\rho T\delta s&=&C_V [\delta T - b_s^{-1} \delta \rho], \\
\delta p&=& \gamma^{-1}c^2\delta\rho+ (1-\gamma^{-1})a_s\delta T, 
\ena  
where $c^2=(\p p/\p \rho)_s$ is the square of 
the sound velocity, 
\be 
\gamma=C_p/C_V  
\en 
is the specific-heat ratio with  
$C_p=\rho T(\p s/\p T)_p$ and  
$C_V=\rho T(\p s/\p T)_\rho$ being  the isobaric and 
constant-volume  
specific heat (per unit volume), respectively.  To avoid cumbersome 
notation, we write  
\be 
a_s= \ppp{p}{T}{s}, \quad 
b_s= \ppp{\rho}{T}{s}=c^{-2}a_s. 
\en 
For low-frequency  sounds, 
the adiabatic relations   $\delta p \cong a_s\delta T$ and 
$\delta \rho \cong b_s\delta T$ should hold.  
We use the following thermodynamic identities \cite{Onukibook},  
\be 
\ppp{p}{T}{\rho}= (1-\gamma^{-1})a_s=  \rho c^2 C_V /Ta_s   . 
\en

Next we consider small hydrodynamic deviations 
by assuming the space-dependence  in 
 the sinusoidal form $\exp(iqx)$.  
From Eqs.(2.3) and (2.5) $\delta\rho$ and $\delta T$ are related by 
\be 
\delta T= \frac{i\omega c^2}{(i\omega+\gamma Dq^2)a_s}\delta\rho, 
\en
where $D=\lambda/C_p$ is the thermal diffusion constant. 
Equations Eq.(2.1)-(2.3) 
give the dispersion equation between $q$ and $\omega$,  
\bea 
&&[\omega^2-(i\omega\nu_{\ell} +\gamma^{-1}c^2) q^2](i\omega+\gamma Dq^2)
\nonumber\\
&&=i\omega c^2 (1-\gamma^{-1})q^2.
\ena 
If we set $q=\omega/c\sqrt{X}$  or $X=(\omega/cq)^2$, 
the dimensionless quantity 
$X$ obeys the quadratic equation, 
\be 
X^2-(1+ \Delta_v+\gamma\Delta_T)X+(1+ \gamma\Delta_v)\Delta_T=0.
\en 
where  we introduce two 
dimensionless  coefficients 
representing the dissipation strength \cite{Moldover}, 
\bea 
\Delta_v&=& i\omega\nu_{\ell}/c^2,\\
\Delta_T&=& i\omega D/c^2.
\ena 
If $\omega>0$, $\Delta_v$ and $\Delta_T$ are purely imaginary. 
The ratio  $\Delta_v/\Delta_T=\nu_\ell/D$ 
grow strongly near the critical point (see Appendix B).

For given $\omega$, Eq.(2.11) or Eq.(2.12) 
 yields  four solutions $q=\pm q_\pm 
=\pm \omega/c\sqrt{X_{\pm}}$,  where $X_+$ and $X_-$ are 
the solutions of Eq.(2.12) written as  \cite{Moldover} 
\bea
X_{\pm}
&=&\frac{1}{2} (1+\Delta_v+\gamma\Delta_T\mp \Xi),\nonumber\\
&=&\frac{2(1+\gamma\Delta_v)\Delta_T}{
1+\Delta_v+\gamma\Delta_T\pm \Xi},
\ena  
where  we define 
\be 
\Xi= [(1+\Delta_v-\gamma\Delta_T)^2+ 4 (\gamma-1)\Delta_T]^{1/2}, 
\en 
with ${\rm Re}\Xi>0$. The second line of Eq.(2.15) follows from 
$X_+X_-= (1+\gamma\Delta_v)\Delta_T$. 
The modes with $q=\pm q_-$ represent the sound, while 
 those  with $q=\pm q_+$  the thermal diffusion. 
We may define $q_-$ and $q_+$ 
such that   ${\rm  Re}(q_-/\omega)>0$ and  ${\rm Im}q_+<0$ 
hold.  It is convenient to introduce $k$ and $\kappa$ by 
\be 
k= q_-=\frac{\omega}{c\sqrt{X_-}} ,\quad 
\kappa= iq_+ = \frac{i\omega}{c\sqrt{X_+}}.
\en
The argument of $X_-$ is in the range $[0,\pi/2]$ for 
$\omega>0$, leading to  ${\rm Im}k<0$, which 
implies  that sound waves  propagating 
in the positive $x$ direction ($\propto e^{-ikx}$) 
are  damped with increasing $x$.  

As $\omega\rightarrow 0$, 
we may treat $\Delta_v$ and $\Delta_T$ 
as small quantities.  To their  first order we find 
$X_+\cong \Delta_T$  and 
$X_- \cong 1+ \Delta_v+ (\gamma-1)\Delta_T$  
so that 
$
\kappa\cong \sqrt{{i\omega}/{D}}$ 
and $
k\cong  {\omega}/{c} - i\Gamma_s {\omega^2}/{2c^3}
$, where  \cite{Landau}   
\be 
\Gamma_s=
(\zeta+4\eta/3)/\rho+ (\gamma-1)D
\en 
is  the   attenuation 
constant in the long wavelength limit.  
We have $|\kappa|\gg |k|$ at low frequencies. 
For example, $|\kappa|\sim 10^5$ cm$^{-1}$ and 
$|k|\sim  10^{-2}$ cm$^{-1}$ for 
$\omega=10^4$ s$^{-1}$,  $D=10^{-6}$ cm$^{2}$s$^{-1}$, 
and $c =10^4$ cm$/$s. In a cell with length $L$, the strength of the 
bulk dissipation of sounds  is represented by the 
damping factor 
$\exp(-\delta_BL)$  with 
\bea 
\delta_B&=&- {\rm Im}k \nonumber\\
&\cong&  \Gamma_s \omega^2 /2\rho c^3, 
\ena
where the second line is the low-frequency expression.  
Mathematically, we may consider the high frequency 
limit $|\Delta_v|\gg 1$ and $|\Delta_T|\gg 1$ 
neglecting the frequency-dependence of the 
 transport coefficients to  derive  
the limiting behavior 
$k \rightarrow (i\omega/\nu_\ell)^{1/2}$  
and $\kappa\rightarrow   (i\omega/\gamma D)^{1/2}$, 
though this limit is unrealistic.  
In this paper, 
we will assume $|\Delta_T|\ll \gamma^{-1}$ in Eq.(2.34), because it  
is satisfied in realistic experimental conditions,  
as will be discussed.

\subsection{Solutions in a finite cell}

We  consider small hydrodynamic 
perturbations behaving as  $e^{\omega t}$   in a fluid 
  in a finite cell with length $L$. 
The  density deviation can be expressed in the following linear 
combination,  
\be 
\delta\rho=  a  e^{-\kappa x} + b   e^{\kappa(x-L)}
+\alpha e^{ikx}+\beta e^{-ikx}. 
\en 
The coefficients  $a$, $b$, $\alpha$, and $\beta$ 
depend on time as  $e^{i\omega t}$.    The first and second terms 
represent the deviations 
 in the thermal diffusion layers.  The  
 thickness of the layers  is 
given by  $1/|\kappa|$, which is 
 assumed to be much shorter than  the  cell length $L$, so  
\be 
|\kappa|\gg 1/L.
\en 
Then the second (first) term is virtually zero 
near $x=0$ ($x=L$).   The third  term in Eq.(2.20) 
represents a sound propagating in the negative $x$ direction 
,  while the fourth  term a sound propagating 
in the  positive $x$ direction.  
 From Eq.(2.1) the  velocity is expressed as 
\be 
v= \frac{i\omega}{\rho\kappa}
 [ a e^{-\kappa x}-b   e^{\kappa(x-L)}]-  
\frac{\omega}{\rho k}[\alpha e^{ikx}-\beta e^{-ikx}].  
\en 
If the boundary walls are  fixed in time,  
we should require 
 $v=0$  at $x=0$ and $L$ to obtain 
\bea 
a &=& (\kappa/ik)(\alpha-\beta),\\
b &=& -(\kappa/ik)(\alpha e^{ikL}-\beta e^{-ikL}) 
\ena
Note that the mass change in the thermal diffusion layers 
is $(a+b)/\kappa$ and that in the interior is 
$\alpha (e^{ikL}-1)/ik +\beta (1-e^{-ikL})/ik$ per unit area. 
From Eqs.(2.23) and (2.24) these two changes cancel, 
ensuring the mass conservation.  
Use of Eq.(2.10) gives the temperature deviation 
in the following linear combination, 
\be 
\delta T= \frac{i\omega}{b_s} 
\bigg [ \frac{a  e^{-\kappa x}
+b e^{\kappa( x-L)}}{i\omega-\gamma D\kappa^2} 
+  \frac{\alpha e^{ikx}+\beta e^{-ikx}}{i\omega+\gamma Dk^2} \bigg]. 
\en 
Let  $\dot{Q}_0$ and  $\dot{Q}_L$ be the 
  heat flux $-\lambda \delta T'$   
 at $x=0$ and $L$, respectively.  
Use of  Eqs.(2.23)-(2.25)  gives  
\be 
\frac{\beta-\alpha}{{\dot Q}_0} 
=\frac{\beta-\alpha e^{2ikL}}{{ e^{ikL}\dot Q}_L} 
= \frac{ b_s(\gamma-1)}{\lambda(1+\gamma\Delta_v)}
 \frac{ik}{k^2+\kappa^2},
\en 
where  $b_s (\gamma-1)/\lambda= \rho/Ta_sD$ 
with the aid of  Eq.(2.9).  The  above 
quantities tend to the constant $\rho/Ta_sc$ in the 
low frequency limit.  We may use Eq.(2.26) when $\dot{Q}_0$ 
is a control parameter or when $\dot{Q}_L$ is  measurable.

The coefficients $a$, $b$, $\alpha$, and $\beta$ 
can be determined if 
we specify the boundary conditions at $x=0$ and $L$. 
Hereafter we assume no temperature discontinuity 
at the boundaries.  In most theoretical calculations  
 the boundary temperatures are fixed, but in some papers 
the bottom heat flux $Q_0$ is fixed \cite{Chiwata}.  
In this paper, 
we consider a more 
realistic boundary condition of the temperature 
accounting for the  thermal conduction 
in the boundary wall regions \cite{Hao,Moldover}. Here  we 
assume that  
$\delta T$ tends to zero in the solid far from the 
boundaries without  heat input.  In  the solid  region 
($x<0$),  the temperature deviation 
then decays  as $\delta T(0)e^{\kappa_w x}$  with    
\be 
\kappa_w=(i\omega C_w/\lambda_w)^{1/2},
\en 
where  $\lambda_w$ and $C_w$ are  the  thermal conductivity 
and the heat capacity (per unit volume) of the solid, 
respectively.   The  $1/|\kappa_w|$ is the  
thickness of the thermal diffusion 
layer  in the solid  and is assumed to 
be shorter than the thickness of the wall. 
Without  temperature discontinuity 
at the boundary, the energy balance  at 
$x=0$  yields     
\bea 
 \delta T'&=& \lambda_w\kappa_w  \delta T/\lambda  \nonumber\\
&=&  a_w  (i\omega/D)^{1/2}  \delta T  , 
\ena 
where $\delta T$ and 
$\delta T'$  are the values  at $x=0$. 
In the second line, 
the coefficient  $a_w$ is the effusivity ratio \cite{Moldover,effusive},   
\be 
a_w=  (C_w\lambda_w/ C_p\lambda)^{1/2}. 
\en 
For CO$_2$ in a Cu cell \cite{Miura} 
we have $a_w= 3\times 10^3\epsilon^{0.92}$.  
The  boundary temperature at $x=0$ is fixed or 
$\delta T(0)=0$ for $a_w\rightarrow\infty$, 
while the  boundary is thermally insulating or 
$(d\delta T/dx)_{x=0}=0$ as $a_w\rightarrow 0$. 
On the other hand, if the other  boundary wall in the region 
$x>L$ is  made of the same material, 
 the boundary condition at $x=L$  reads  
\be 
 \delta T'
= - a_w  (i\omega/D)^{1/2}  \delta T  , 
\en
with the same $a_w$ as in Eq.(2.28), 
where $\delta T$ and 
$\delta T'$  are the values  at $x=L$.

The boundary 
conditions at $x=0$ give 
Eqs.(2.23) and (2.28), from which 
we may readily calculate  the 
reflection factor $Z\equiv \beta/\alpha$ 
between the outgoing and incoming sound waves. 
It is convenient to introduce the combination,    
\be 
W=  \frac{\alpha-\beta}{\alpha+\beta} =\frac{1-Z}{1+Z},
\en  
because $W$ is a small quantity in our system.
Some calculations yield   a general expression,     
\be 
W= \frac{-ik (i\omega-\gamma D\kappa^2)/\kappa}{{i\omega+\gamma DJ^2} 
+ \sqrt{i\omega D} (\kappa^2+k^2)/a_w \kappa},
\en
in terms of $k$ and $\kappa$. In the case of 
a thermally insulating boundary, 
 we have $W=0$ and $Z=1$ 
by setting  $a_w\rightarrow 0$ in Eq.(2.32).  
The interaction of sounds and the 
boundary wall is characterized by 
 $Z$ or $W$, where the wall properties appear 
only through the effusivity ratio $a_w$ 
and the system length $L$ does not appear.

\subsection{Adiabatic condition  in the interior }

We will clarify an  upper bound 
of the frequency, below which 
the sound motions  in the interior are 
{\it adiabatic} or without entropy deviations. 
Under this adiabatic condition, 
the results from the linear hydrodynamic 
equations can be much simplified.

Far from the boundary walls or outside the thermal diffusion layers,
we may neglect  the localized modes  
to obtain the interior hydrodynamic deviations. 
From Eqs.(2.5),  (2.6), and  (2.25), those  of the 
density, temperature, and pressure   are  related by  
\bea 
\delta \rho 
&=& [1+\gamma D k^2/i\omega] b_{s}\delta T,\nonumber\\
\delta p
&=& 
[1+ D k^2/i\omega] a_{s}\delta T.
\ena 
Here $x$ and $L-x$ are much longer than $ 1/|\kappa|$.
The second terms in the brackets  arise 
 from a small entropy deviation  in
the interior.  Since $\gamma>1$, 
the usual adiabatic relations  
hold in the interior  under the condition,  
\be 
|\gamma D k^2/i\omega|   
\sim \gamma |\Delta_T| \ll 1 \quad {\rm or} \quad 
\omega \ll c^2/\gamma D,  
\en 
where $\Delta_T$ is defined by Eq.(2.14). 
This  condition 
is well satisfied 
in the usual hydrodynamic processes. Even near the critical point,  
the time $t_{ad}\equiv \gamma D/c^2$  remains very short. 
For example,  $t_{ad}= 7.6\times 10^{-14}\epsilon^{-0.62}$sec 
for CO$_2$.

Under Eq.(2.34)  we have $\Xi\cong 
1+\Delta_v$ so that  $X_+$ and $X_-$ in Eq.(2.15) 
are approximated as 
\be 
X_+= \frac{1+\gamma\Delta_v}{1+\Delta_v}\Delta_T , 
\quad  X_-= 1+\Delta_v.  
\en 
The wave numbers $k$ and $\kappa$ are expressed as 
\be 
\kappa =\bigg(\frac{i\omega}{D}\bigg)^{1/2}
\bigg(\frac{1+\Delta_v}{1+\gamma\Delta_v}
\bigg)^{1/2}, 
\quad   
 k= \frac{\omega/c}{\sqrt{1+\Delta_v}} .  
\en 
We retain $\Delta_v$, 
since  it  becomes appreciable near 
the critical point because of the strong critical 
divergence of $\zeta$. 
As will be discussed in Appendix B, $\zeta \cong \rho c^2 
R_B t_\xi$  for $\omega t_\xi <1$, where 
$R_B \cong 0.03$ is a universal number 
and $t_\xi=\xi^2/D$ is the characteristic 
time of the critical fluctuations
 with $\xi$ being the correlation length. Carl$\grave{\rm{e}}$s 
 found  the dependence of $\kappa$ 
on the singular combination $\gamma \zeta$ as in Eq.(2.36) 
\cite{Carles0,Carles}.   By setting 
$\gamma \Delta_v=i\omega t_B$, 
we introduce  a new characteristic time $t_B$ as     
\be 
t_B= \gamma\zeta/\rho c^2=R_B  \gamma t_\xi.  
\en 
Then  $t_B\gg t_\xi$  
once $R_B\gamma\gg 1$.  
For CO$_2$,  $t_B= 1.9\times 10^{-15}\epsilon^{-3.0}$sec. 
See   Table 1 and Fig. 1 for the 
 characteristic  times with $L=1$cm, 
where $t_B$  exceeds  the acoustic  time $L/c$ 
for $\epsilon<3\times 10^{-4}$  and the modified piston 
time $t_1'$ (to be introduced in Eq.(3.16)) for 
$\epsilon<3\times 10^{-5}$.  
There can be a sizable frequency range with 
$t_B^{-1} < \omega <t_\xi^{-1}$ at small $\epsilon$, where 
$\kappa$ becomes independent of $\omega$ as 
\be 
\kappa\cong (\rho c^2/\gamma D\zeta)^{1/2}
 \cong (R_B\gamma)^{-1/2}\xi^{-1}. 
\en 
The thickness  of the 
thermal diffusion layer   $1/|\kappa|$ remains longer than $\xi$ 
by $(R_B\gamma)^{1/2}$. 
 Also from the expression of $k$ in Eq.(2.36) 
we write the sound dispersion relation 
as $k= \omega/c^*(\omega)$, where 
we  define the complex sound velocity\cite{Onukibook},  
\be
c^*(\omega)= c\sqrt{1+\Delta_v}, 
\en 
whose critical behavior will be discussed in Appendix B.

\begin{table}
\caption{Parameters of CO$_2$ in a Cu cell with 
$L=1$ cm for 
 $\epsilon=10^{-3}$ (first line), 
 $10^{-4}$ (second line),  and 
 $10^{-5}$ (third line). Times are  in sec.}

\begin{tabular}{cccccccc}
 $\gamma$& $a_w$ &
 $t_\xi\times 10^6$ & $t_B\times 
10^6$& ${L}/{c}\times 10^4$& $t_1'$
 & $t_2\times 10^8$  \\ 
\hline 
  260&5.0 &  0.24 & 1.9&0.71&1.0 &0.12 \\
\hline
 3600&0.63
 & 18 & 1900&0.83&0.1&1.7 \\
\hline
 $5\times 10^4$ & 0.075
 & 1300 & $1.7\times 10^6$ & 0.98&0.08&3.0 \\
\hline
\end{tabular}
\end{table}

\begin{figure}[th]
\includegraphics[scale=0.46]{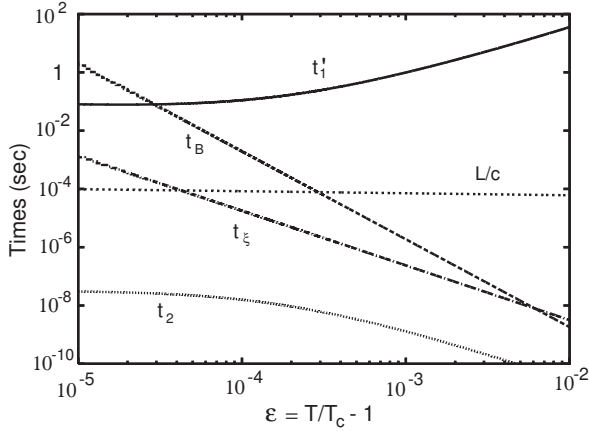}
\caption{Characteristic times 
$t_\xi$ in Eq.(B3), $t_B$ in Eq.(2.37), $t_2$ in Eq.(2.43), 
$t_1'$ in Eq.(3.16), and 
$L/c$ vs $\epsilon=T/T_c-1$ 
for  CO$_2$ in a Cu cell with 
$L=1$cm.
 }
\end{figure}

\begin{figure}[th]
\includegraphics[scale=0.31]{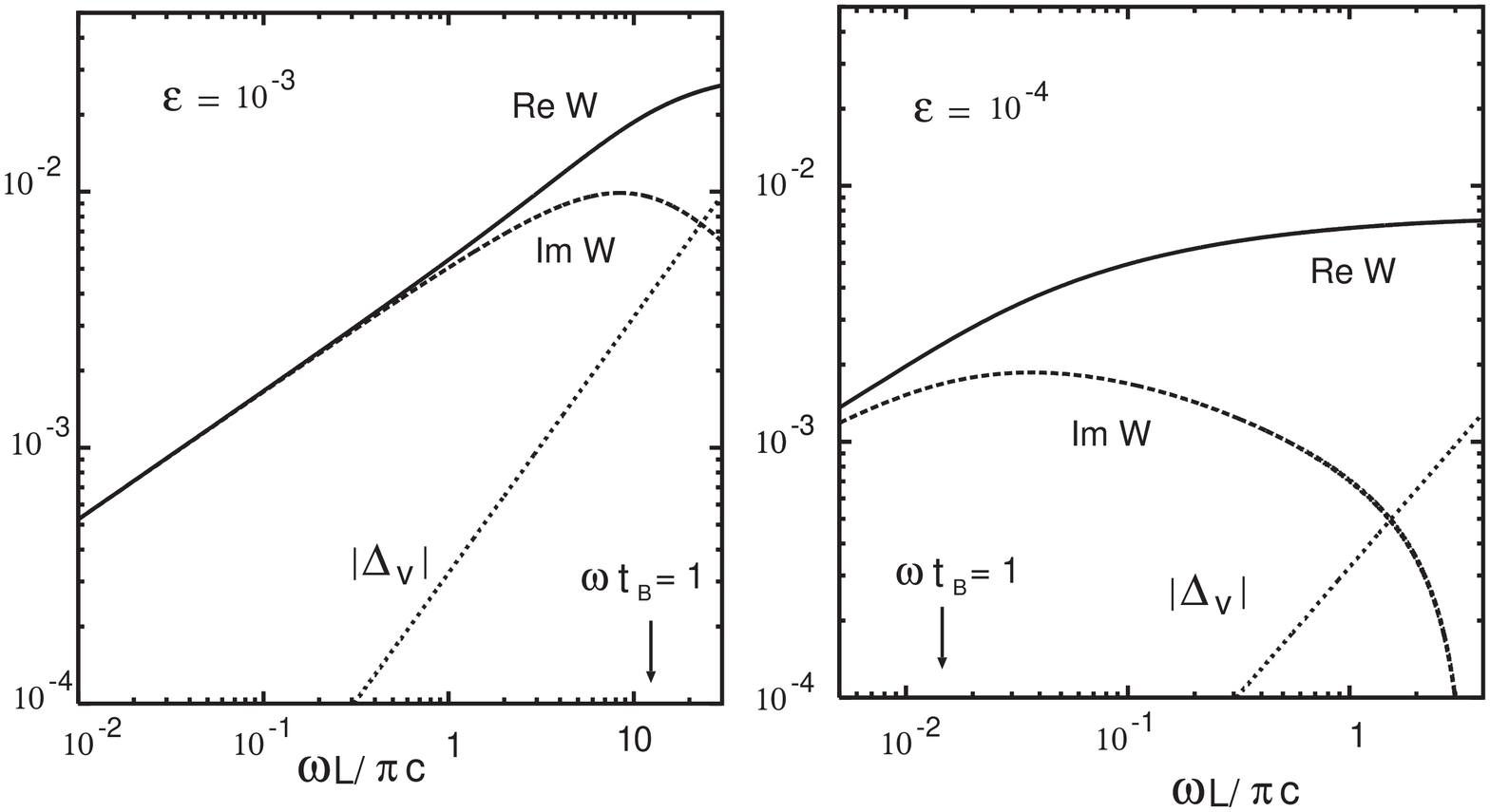}
\caption{${\rm Re}W$, ${\rm Im}W$, 
and $|\Delta_v|= \omega\zeta/\rho c^2$ 
vs $\omega L/\pi c$ 
at   $\epsilon=10^{-3}$ (left) 
and $10^{-4}$ (right) for 
CO$_2$ in a Cu cell with 
$L=1$cm.  As $\omega$ exceeds $t_B^{-1}$,  
 ${\rm Re}W$ tends to saturate and 
 ${\rm Im}W$  decreases, due to the 
growing bulk viscosity and the decreasing effusivity ratio. 
}
\end{figure}


As $\epsilon$ is decreased, 
we first encounter the regime where 
  $W$ grows  but  $a_w\gg 1$ and 
$\omega t_B\ll 1$ still hold.  However, 
the critical growth of  $W$ 
is eventually suppressed by 
 the growing    $a_w^{-1}$ 
and  $\zeta$. If we use   Eqs.(2.35) and (2.36) under Eq.(2.34), 
we  approximate  $W$ 
in  Eq.(2.32) as   
\be 
W=\frac{(\gamma-1)\sqrt{\Delta_T}}{(1+\Delta_v)X_v},
\en
where  we define 
\be
X_v= \sqrt{1+\gamma\Delta_v}+a_w^{-1}\sqrt{1+\Delta_v}.
\en 
The limiting behaviors of $X_v$ are as follows: 
$X_v \cong 1+a_w^{-1}$ for $\omega\ll t_B^{-1}$ 
and $X_v \cong (i\omega t_B)^{1/2}$  for $\omega\gg t_B^{-1}
(1+a_w^{-1})^2$.

For $\omega\ll t_B^{-1}$,  it follows  the 
  classical expression valid far from the critical point,   
\be 
W =(\gamma-1)\sqrt{\Delta_T}/(1+a_w^{-1})=  \sqrt{it_2\omega},
\en  
which is the result without the viscous   
effect and under the isothermal boundary condition. 
We may introduce  a  characteristic time  $t_2$ 
defined by 
\be 
t_2= [a_w/(1+a_w)]^2(\gamma-1)^2D/c^2,
\en 
which includes  the effect of the  heat conduction 
in the wall.   As shown in  Table 1 and Fig. 1, 
$t_2$ is very short even compared with $t_\xi$.
In the literature (see Section 77 of Ref.\cite{Landau}),  
it is argued that the amplitude  
of a plane wave sound is decreased 
by the factor  $(\gamma-1)\sqrt{2D\omega}/c$ 
upon reflection at 
an isothermal boundary wall. This factor 
 is obviously equal to $1-|Z|\cong 2{\rm Re}W$ 
if use is made of Eq.(2.42).

In Fig. 2, ${\rm Re}W$  and  ${\rm Im}W$  
are displayed as functions of  $\omega$ 
at   $\epsilon=10^{-3}$ 
and $10^{-4}$. 
While  $\omega t_B<1$, they increase 
with increasing $\omega$ obeying Eq.(2.42).  
After  $\omega$ exceeds $t_B^{-1}$, 
$\sqrt{1+\gamma \Delta_v}$ becomes 
$\sqrt{\gamma \Delta_v}$  in Eq.(2.41); then,  
 ${\rm Re}W$ tends to saturate and 
 ${\rm Im}W$  decreases. In fact, for $\omega\gg t_B^{-1}
(1+a_w^{-1})^2$,  we have $X_v \cong \sqrt{\gamma \Delta_v}$ and
\be 
W \cong   (\gamma D/\nu_\ell)^{1/2}  \cong W_0 \epsilon^{0.64},
\en   
where $W_0\cong 2.1$ for CO$_2$. 
Also as a function of $\epsilon$,    
${\rm Re}W$ exhibits a maximum 
around the reduced temperature  
at  which $t_B\sim \omega^{-1}$, as   will be 
shown  in Fig. 3.
Growing  $a_w^{-1} (\propto \epsilon^{-1.14})$ 
 further serves to decrease $W$. Thus, 
even close to the critical 
point,  we find $|W|\ll 1$ and  
\be 
Z=1-2W+\cdots . 
\en

\subsection{Hydrodynamic variables in the adiabatic condition}

Under Eq.(2.34) we obtain simple 
expressions of  the deviations of the temperature, 
the pressure, and the entropy including $\Delta_v$. 
From Eqs.(2.5), (2.6), (2.9), and (2.20) we find  
\bea 
\delta T &=& \ppp{T}{\rho}{p} ({1+\gamma \Delta_v})
[\delta \rho]_b + {b_s}^{-1} [\delta \rho]_{in} \\
\delta s &=& \ppp{s}{\rho}{p}(1+\Delta_v) [\delta \rho]_b ,\\
 {\delta p} &=& -c^2  \Delta_v
[\delta \rho]_b+c^2[\delta \rho]_{in}, 
\ena 
where 
$[\delta \rho]_b= a e^{-\kappa x}+b   e^{\kappa(x-L)}$ 
is the density deviation localized near the boundaries 
and 
$[\delta \rho]_{in}= \alpha e^{ikx}+\beta e^{-ikx}$ 
is the interior density deviation. 
In deriving Eq.(2.46) use has been made of 
the thermodynamic relation $(1-\gamma)b_s= (\p \rho/\p T)_p$. 
Under Eq.(2.34), the entropy deviation $\delta s$ is 
 localized near the boundaries, while 
the localized part of the pressure deviation $\delta p$ 
is nonvanishing to satisfy 
$i\omega\rho v= 
-\delta p'+\rho\nu_\ell v''= -\delta p' 
-i\omega\nu_\ell\delta \rho' \rightarrow 0$ 
as $x\rightarrow 0$ and $L$ in Eq.(2.2).

We examine the  deviations  close to the boundary 
at $x=0$ by asssuming Eqs.(2.23) and (2.28) and setting  
$Z\alpha=\beta$.  
In this case the density ratio 
$[\delta \rho]_b/[\delta \rho]_{in}$ 
tends to  
$
{(\gamma-1)}/{X_v\sqrt{1+\gamma\Delta_v}}$ 
as $x\rightarrow 0$, so that    
\bea 
b_s {\delta T}&=& ({\alpha+\beta}) 
\bigg[1- 
\frac{1}{X_v} {\sqrt{1+\gamma\Delta_v}} 
e^{-\kappa x} \bigg],
\\ 
\frac{\delta p}{c^2} &=& (\alpha+\beta)\bigg[1-
\frac{(\gamma-1)\Delta_v}{X_v\sqrt{1+\gamma\Delta_v}} 
e^{-\kappa x}\bigg] ,
\ena 
where $0<x\ll 1/|k|$.  
Note that the second terms 
 in the brackets in Eqs.(2.49) and (2.50) tend  to unity 
for $\omega t_B \gg (1+a_w^{-1})^{-2}$, which can be 
the case of very large $\zeta$.  
In the original  work \cite{Ferrell}, 
the pressure homogeneity  
and  the isobaric relations  among 
the hydrodynamic variables were assumed 
in the thermal diffusion layers. 
We recognize that the pressure homogeneity and 
the isobaric condition  hold  only in 
 the low frequency limit 
$\omega\ll t_B^{-1}(1+a_w^{-1})^{-2}$.

\section{Applications} 
\setcounter{equation}{0}

\subsection{Acoustic modes in a cell}

   Gillis {\it et al.}
\cite{Moldover}  
calculated the  acoustic eigenmodes for their 
experimental geometry, taking  into account 
the growing  $a_w^{-1}$ and $\zeta$. 
In the following,  we will  present 
a simpler version  in a one-dimensional cell,
$0<x<L$,  taking  into account  
these  two ingredients. 
In this case  $\omega$ is treated as 
one of the eigenvalues 
and is complex,  while 
we have assumed $\omega>0$ in the previous 
section.   Then  $\omega$ should 
have a positive imaginary part 
for the stability of the system. 
Here $\sqrt{i\omega}= (1+i)\sqrt{\omega/2}$ for 
$\omega>0$,  while    
  $\sqrt{i\omega}= (1-i)\sqrt{|\omega|/2}$ for  $\omega<0$. 
The latter follows from the requirement that the 
 real part of $\kappa$ in Eq.(2.17) should be positive.   
For general complex $\omega$, all the quantities 
introduced so far should be   functions  
of $\omega$ analytic for ${\rm Re}(i\omega)>0$ 
or  for ${\rm Im}\omega<0$.   Therefore,  
$X_\pm(-\omega) = X_\pm (\omega^*)^*$ and 
\be 
Z(-\omega^*)=Z(\omega)^*, \quad W(-\omega^*)=W(\omega)^*, 
\en 
where  $\omega^*$ is  the complex 
conjugate of $\omega$.

  Under the boundary conditions (2.28) and (2.30), 
the interior density deviation is expressed  as 
\bea 
\delta \rho  &=&  \alpha (e^{ikx}+Ze^{-ikx}),\nonumber\\ 
&=&  \alpha' (e^{ik(L-x)}+Ze^{-ik(L-x)}), 
\ena 
in terms of $Z=\beta/\alpha$.  
 The  first and second lines  
follow from Eqs.(2.28) and (2.30), respectively, and 
should coincide 
so that 
$\alpha'e^{ikL}= \beta$ and $\alpha'Ze^{-ikL}= \alpha$, leading to      
$Ze^{-ikL}= Z^{-1}e^{ikL}$.
We now find   the condition of the eigenmodes, 
\be 
Z= \pm e^{ikL}, 
\en 
where $+$ corresponds to even modes and 
$-$  to odd modes. Namely, the density and temperature 
deviations are even (odd) functions of $x-L/2$ for the the even (odd) modes. 
For $W= (1-Z)/(1+Z)$ calculated in Eq.(2.32) or Eq.(2.40),   we obtain 
\bea 
W &=& -i \tan  (kL/2)\quad ({\rm even~modes})\nonumber\\ 
&=& i \cot  (kL/2)\qquad ({\rm odd~modes}).
\ena

\begin{figure}[t]
\includegraphics[scale=0.31]{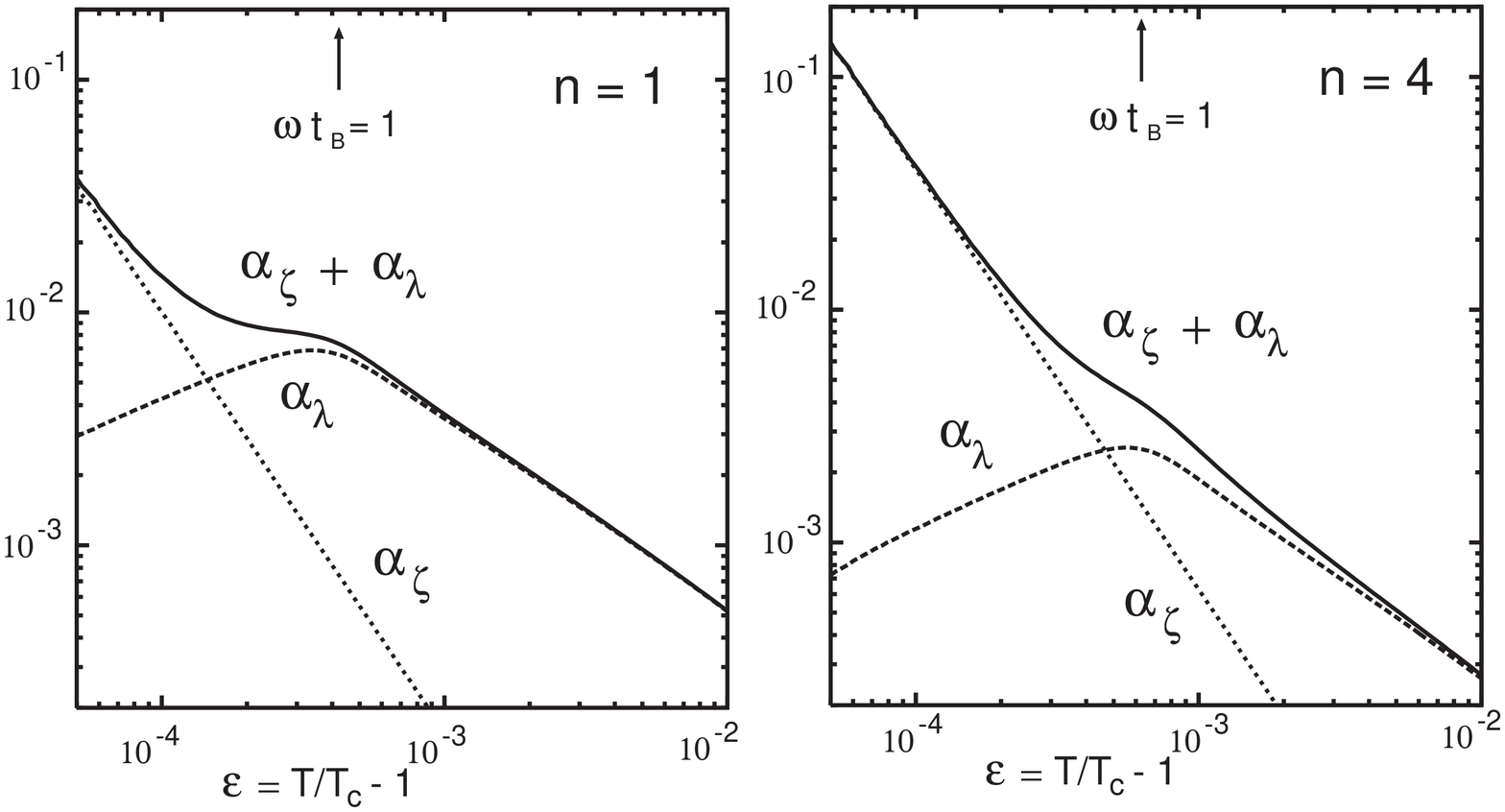}
\caption{Normalized damping 
constants 
$\alpha_\lambda$, $\alpha_\zeta$, 
and $\alpha_\lambda+ \alpha_\zeta$ 
defined by (3.5) and (3.6) vs $\epsilon 
=T/T_c-1$ for $n=1$ (left) 
and $n=4$ (right). The resonant frequencies are 
close to $\omega=n\pi c/L \sim n\times 4\times 10^4$sec$^{-1}$, 
which are exceeded by  $t_B^{-1}$ close to the critical point as 
marked by the arrows.  
}
\end{figure}

Since  $W$ is small,  the eigenfrequencies 
$\omega_n$ $(n=1,2,3\cdots)$ are nearly equal 
to $n\pi c^*/L$ where $c^*$ is defined by Eq.(2.39).  
In  the case $|W|\ll 1$   the leading correction 
from  $W$  can be written as   
\bea 
\omega_n &=& ( n\pi+2iW+\cdots)c^*/L,\nonumber\\
&=& [1+ i\alpha_\lambda+{i}\alpha_\zeta 
+\cdots]{\rm Re}\omega_n,
\ena 
where $n=2,4, \cdots$ for the even modes  and  
$n=1,3, \cdots$ for the odd modes. 
We  assume  small bulk damping 
$|\Delta_v|\ll 1$ and define 
\be 
\alpha_\lambda= \frac{2}{n\pi}{\rm Re}W, \quad 
\alpha_\zeta=\frac{1}{2}  |\Delta_v|
=n\pi \frac{\zeta}{2\rho cL},
\en 
where $\alpha_\lambda$  represents the boundary damping 
and $\alpha_\zeta$  the bulk damping.  
The resonance quality factor $Q^{-1}$ 
\cite{Moldover} is equal to $2(\alpha_\lambda+\alpha_\zeta)$ 
in our notation. The resonance frequency including the 
shift is given by the real part, 
\be 
{\rm Re}\omega_n= ( n\pi- 2{\rm Im}W+\cdots) {\rm Re}c^*/L.
\en  
The frequency   $\omega$ in  $c^*$ 
and $W$ may be equated with ${\rm Re}\omega_n  
\cong n\pi c/L$.

In Fig. 3, we show $\alpha_\lambda$, 
$\alpha_\zeta$, and the sum $\alpha_\lambda+\alpha_\zeta$ 
as functions of $\epsilon$ 
in the regime  $\omega t_\xi<1$ 
for the odd mode of $n=1$ and the even mode of $n=4$ 
for CO$_2$ in a Cu cell with $L=1$cm. 
We notice the following. 
(i) For such long wavelength sounds, the boundary damping 
is relevant far from the critical point,  but 
the bulk damping eventually dominates 
 close to the critical point. 
(ii) In accord with the discussion around Eq.(2.44), 
$\alpha_\lambda$ decreases on approaching the criticality 
in the region of $t_B>\omega^{-1}\sim L/n\pi c$, 
with  a maximum  at $ \omega t_B\sim 1$.  
As a result,  the curve of the sum $\alpha_\lambda+\alpha_\zeta$ 
is flattened considerably around $t_B\sim\omega^{-1}$.

These theoretical results are consistent with 
 the experimental data  by 
Gillis  {\it et al.}\cite{Moldover}. 
They performed the resonance experiment 
over a wide range of  $\omega t_\xi$ (up to about 200)  
to measure the frequency-dependent bulk viscosity. 
In agreement with  the theory \cite{Onukibook,Ferrell}, 
 $\alpha_\zeta$ or $\omega\zeta/\rho c^2$ 
became independent of 
$\epsilon$ in the high-frequency regime 
 $\omega t_\xi>1$ (see Eq.(B5) in Appendix B).

\subsection{Periodic perturbations }

Periodic perturbations 
may be  applied to a  fluid in a cell in various manners.  
Resonance can occur  when the frequency $\omega$ is close to 
${\rm Re}\omega_n$. 
It is sharp for small 
  ${\rm Im} \omega_n$. 
We will give three  boundary conditions  
at $x=0$ leading to resonance. 
We  assume the boundary condition  Eq.(2.30) 
at $x=L$. Then  use of   Eq.(3.2)  yields  
$\alpha= \beta Ze^{-2ikL}$. 
The interior density deviation  is  of the form, 
\be 
\delta \rho=\beta e^{-ikx}+ \beta Ze^{ikx-2ikL}, 
\en    
where the  term proportional to $Z$   arises   from 
the reflection at $x=L$.  

The bulk damping of 
the reflected waves is represented by $|e^{-ikL}|=e^{-\delta_BL}$. 
From  Eq.(2.19) $\delta_B$ is expressed as  
\be 
\delta_B= A_B (\omega /\pi c)^2,
\en  
in the low frequency regime $\omega t_\xi <1$. 
We find 
$A_B= 0.5\times 10^{-3}$cm   
and $0.03$cm   at  
$\epsilon=10^{-3}$ 
and $10^{-4}$, respectively, for CO$_2$. 
In the relatively high frequency range 
$\omega  > (2A_B)^{-1/2}\pi c/L$,  
the factor  $e^{-2ikL} (\propto e^{-2\delta_BL})$ becomes 
negligibly small. Then, near the boundary, 
 $\delta\rho$ consists of the 
outgoing wave only, resulting in no   resonance. 
On the other hand, in the high  frequency regime $\omega t_\xi >1$,  
Eq.(B6) gives 
\be 
\delta_B \cong \omega {\rm Im}\Delta_v
/2c \cong 0.27 \omega /\pi c, 
\en 
which means that a sound emitted at $x=0$ 
reaches the other end 
with the damping factor $e^{-0.27}$ 
for the first resonance frequency  $\omega\cong \pi c/L$.

\subsubsection{Temperature oscillation}

\begin{figure}[th]
\includegraphics[scale=0.37]{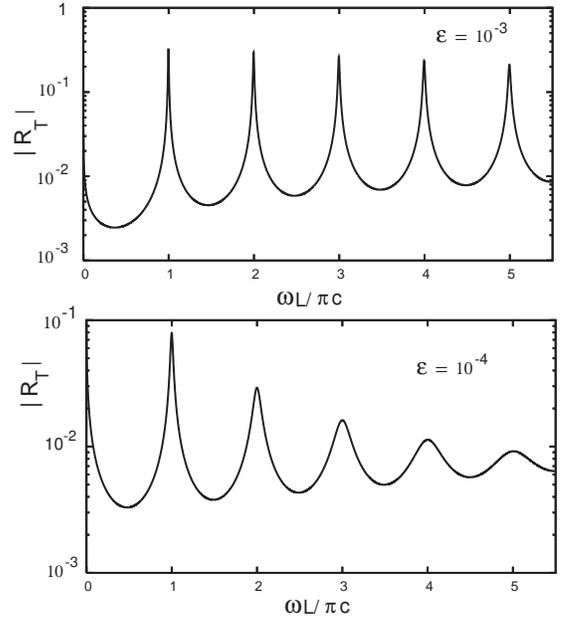}
\caption{Absolute value of the response 
function $R_T(\omega)$ in Eq.(3.12) vs $\omega L/\pi c$ 
on a semi-logarithmic scale for $\epsilon=10^{-3}$ (upper panel) 
and $10^{-4}$ (lower panel) applicable    for 
CO$_2$ in a Cu cell with 
$L=1$ cm.  
 }
\end{figure}

In the first example,  
the temperature in the wall region $x<0$ is oscillated, 
while the boundary walls are 
mechanically fixed. More precisely, we  require 
$\delta T \rightarrow 
T_w \propto e^{i\omega t}$ as  
$x\rightarrow -\infty$; then,  
 $\delta T (x)= e^{\kappa_w x}[ \delta T(0)-T_w] 
+T_w$ in the region $x<0$.    The thermal 
boundary condition at $x=0$ 
is then given by  
\be 
 \delta T'=  a_w  (i\omega/D)^{1/2}  (\delta T- T_w) ,  
\en
as a generalization of Eq.(2.28).  
Some calculations using Eqs.(2.23) and (2.25) give  
the response function defined by 
$R_T \equiv {\beta}/{b_s T_w}$ in the form,  
\be
R_T = \frac{1}{2} 
(1+ \gamma Dk^2/i\omega)\frac{1-Z}{1-Z^2e^{-2ikL}}. 
\en 
Notice that $R_T$ diverges  
as $R_T\cong {cW}/{2iL(\omega-\omega_n)}$ 
for $\omega=\omega_n$ ($n=1,2,\cdots)$ in the complex 
$\omega$ plane from Eq.(3.3). 
Under the adiabatic condition Eq.(2.34),  the  
 interior  temperature deviation is  
expressed   as  
\be 
\delta T = b_s^{-1} {\delta \rho}=
( e^{-ikx}+  Ze^{ikx-2ikL})R_T T_w.
\en     
Furthermore,  neglecting $\gamma Dk^2/i\omega$ 
in Eq.(3.12) and using  $|W|\ll 1$ (see Fig. 2), we obtain 
\be 
R_T\cong W/[1-(1-4W)e^{-2ikL}]. 
\en

In Fig. 4, 
we plot the absolute value $|R_T|$ calculated from 
 Eq.(3.12) vs the normalized frequency  
$\omega L/\pi c$  at   
$\epsilon=10^{-3}$ and $10^{-4}$, 
using the data  for CO$_2$ in a  Cu cell with $L=1$cm \cite{Miura}. 
It exhibits peaks at $\omega \cong n\pi c/L$ as expected, 
but its peak heights do not exceeds $1/2$ due to 
the small factor $1-Z\cong 2W$ in the numerator in Eq.(3.12) 
 As discussed below Eq.(3.9), 
the  resonant peaks should disappear 
for $\omega L/\pi c> (2A_B)^{-1/2}$, 
where we may neglect  $e^{-2ikL}$ in $R_T$   to obtain 
$
R_T \cong W.  
$ 
These results are in accord with Fig. 4,  since  
$(2A_B)^{-1/2}\cong $30 and 4 for 
$\epsilon=10^{-3}$ and $10^{-4}$, respectively.

 In the low frequency case 
$\omega \ll c/L$,  the interior deviations become nearly 
homogeneous.  Figure 1 indicates that $t_B$ can much 
exceed $L/c$ very close to the critical  point, while 
 $|\Delta_v|\ll 1$ holds. Thus, retaining $\gamma \Delta_v$, we 
   set $e^{ikL}\cong 1+ikL$ and   $1+\Delta_v \cong 1$ 
 and  use Eqs.(2.40) and Eq.(3.12) to find 
\be 
R_T\cong \frac{1}{4}[{\sqrt{i\omega t_1}}X_v+1]^{-1}, 
\en 
where $t_1$ is defined by Eq.(1.1) and $X_v$ by Eq.(2.41).
If $\omega \ll t_B^{-1}$, we further have 
$R_T \cong ({\sqrt{i\omega t_1'}}+1)^{-1}/4$ with  
\be 
t_1'= (1+a_w^{-1})^2t_1 = (1+a_w^{-1})^2 L^2/4(\gamma-1)^2D,  
\en  
which is related to $t_2$ in Eq.(2.43) by 
$t_1' t_2 =L^2/4c^2$.  The $t_1'$  
first decreases   as $t_1\sim
\epsilon^{2.26}$ for $a_w\gg 1$ 
but finally weakly increases  
 as $t_1a_w^2 \cong \epsilon^{-0.22}$  for $a_w \ll 1$.  See  Fig. 1 
for the curve of $t_1'$.    
As will be discussed in Subsection III C, 
 $t_1'$ is the piston time including the effect 
of the wall heat conduction \cite{Hao}.

We need to know when  
$|\sqrt{i\omega t_1}{X_v}|\gg  1$ holds. 
It holds for  
$\omega \gg  1/t_1'$ under the condition, 
\be 
(1+a_w^{-1})^{-2} t_B  /t_1' =(1+a_w^{-1})^{-4}  t_B/t_1   \ll 1. 
\en 
If $a_w<1$ for CO$_2$ in a Cu cell, 
the above condition becomes  
 $a_w^4t_B/t_1= 2\times 10^{-6}\epsilon^{-0.97}/L^2\ll 1$ 
with $L$ in cm , which  is well satisfied 
for $\epsilon \gg 10^{-6}$ with $L=1$cm. 
 If $\omega t_1'\ll 1$ under Eq.(3.17), we find 
\be 
\delta T \cong T_w/2,
\en 
in the interior.  Note that the reverse condition of Eq.(3.17), 
  $a_w^4t_B/t_1 >1$,  holds extremely close to the critical point, 
where  $|\sqrt{i\omega t_1}{X_v}|> 1$ and 
$R_T \cong 1/4i\omega\sqrt{t_1t_B}$ 
are obtained  for 
$\omega \gg  (t_1t_B)^{-1/2}$.  
See the discussion below Eq.(3.32) 
for the relaxation behavior in this ultimate regime.

Zhong {\it et al.} \cite{Kogan} measured a  density 
change induced by boundary temperature 
oscillation in  near-critical $^3$He, 
where the frequency was very low ($\omega/2\pi <2$Hz) 
and the bulk viscosity was not important. 
However, they could measure in-phase and out-of-phase 
response in agreement with the original theory \cite{Ferrell}.

\subsubsection{Mechanical oscillation}

In the second example, 
the boundary wall at $x=0$ is mechanically 
oscillated  without  heat input from outside.  
This is the case in the usual acoustic experiments  using a piezoelectric 
transducer\cite{Moldover}. 
Let $u_w (\propto e^{i\omega t})$ be 
the applied displacement amplitude; then, 
\be 
v= i\omega u_w 
\en 
 at $x=0$ in Eq.(2.12).   Assuming Eq.(2.30) 
and using Eq.(3.8) we obtain 
\bea
\beta&=& -\bigg[1+\frac{\kappa}{a_w} 
\sqrt{\frac{D}{i\omega}}\bigg]
\frac{\kappa R_T}{1-\gamma D\kappa^2/i\omega}
\rho{u_w} \nonumber\\
&=& ({\gamma-1})^{-1}\sqrt{{i\omega}/{D}} 
X_vR_T \rho{u_w}
\ena  
where the first line is general and 
the second line is the approximation 
under the adiabatic condition Eq.(2.34). 
Since the response is proportional to $R_T$, 
resonance occurs as in the previous case of 
temperature oscillation.

In the low frequency case  $\omega \ll c/L$,  
the interior density change is nearly homogeneous and 
\be 
\delta\rho \cong 2
\beta \cong 
\bigg [1- \frac{1}{\sqrt{i\omega t_1}X_v+1}\bigg ]
\frac{ \rho u_w}{L} ,
\en 
which is the counterpart of Eq.(3.14). 
As discussed below Eq.Eq.(3.16), 
$\sqrt{i\omega t_1}X_v$ is  large 
in the  relatively high-frequency 
range  $\omega \gg  1/t_1'$ under Eq.(3.17). Thus 
 the interior density deviation behaves as  
\bea 
\delta\rho &\cong&  \rho u_w/L \quad  (1/t_1'\ll \omega \ll   c/L) \nonumber\\
 &\cong&  \sqrt{i\omega t_1}X_v 
\rho u_w/L  \quad ( \omega \ll   1/t_1'), 
\ena 
under Eq.(3.17). 
The volume  change  mostly 
occurs in the bulk region for $1/t_1'\ll \omega \ll   c/L$ 
and in  the thermal diffusion layers 
for  $\omega t_1'\ll 1$.

\subsubsection{Heat flux oscillation}

\begin{figure}[th]
\includegraphics[scale=0.37]{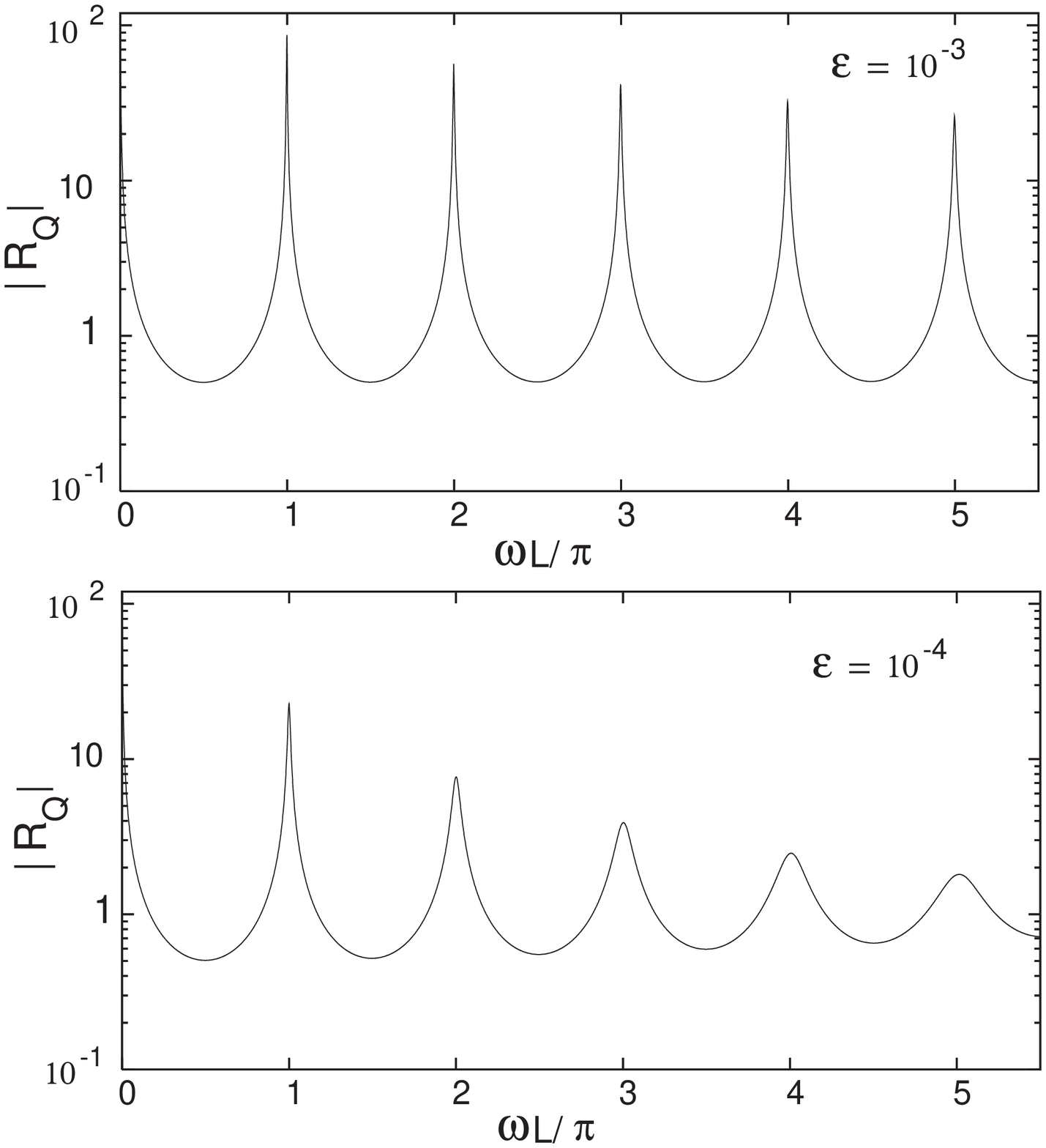}
\caption{Absolute value of the response 
function $R_Q(\omega)$ in Eq.(3.24)   
vs $\omega L/\pi c$ 
on a semi-logarithmic scale for $\epsilon=10^{-3}$ (upper panel) 
and $10^{-4}$ (lower panel) applicable    for 
CO$_2$ in a Cu cell with 
$L=1$ cm.  
}
\end{figure}

In the third example, we    apply a 
 heat flux ${\dot Q}_0=-\lambda (dT/dx)_{x=0} 
 \propto e^{i\omega t}$ at $x=0$ 
assuming the boundary condition (2.30). 
It is convenient to 
 introduce a dimensionless response function 
$R_Q$ by 
\be 
\beta = \frac{\rho}{cT}\ppp{T}{p}{s} R_Q \dot{Q}_0 
\en  
Then  Eqs.(2.26) and Eq.(3.8) give  
\bea 
R_Q &=& \frac{ick/D}{(1+\gamma \Delta_v)(k^2+\kappa^2)(1-Ze^{-2ikL})} 
\nonumber\\
 &=& \frac{1}{(1+ \Delta_v)^{3/2}(1-Ze^{-2ikL})}, 
\ena 
where the first line is general and 
 the second line holds under the adiabatic condition 
(2.34).  In the complex $\omega$ plane, 
$R_Q$ has poles $\omega_n'$, which are equal to 
$\omega_{2n}$ in Eq.(3.5) with system length changed to $2L$. 
Thus $R_Q$ grows for $\omega \cong n \pi c/L$ 
for  $\omega < (2A_B)^{-1/2}\pi c/L$. 
In Fig. 5, we plot the absolute value 
$|R_Q|$ as a function of 
$\omega L/\pi c$ for $\epsilon=10^{-3}$ and $10^{-4}$. 
We can see that  $|R_Q|$ 
is larger than  $|R_T|$ in Fig. 4 roughly 
by two orders of magnitude.

The behavior of $R_Q$ in the low frequency range  
$\omega \ll c/L$  is very different from that of $R_T$, however. 
From the second line of Eq.(3.24)   we have   
$
R_Q\cong 1/(2ikL+2W) 
$   
to obtain the counterpart of Eq.(3.15),   
\be 
R_Q 
\cong \frac{X_v}{2\sqrt{i\omega t_2}}
[1+ 2\sqrt{i\omega t_1}X_v]^{-1}. 
\en
Under Eq.(3.17)  we find that 
$R_Q \cong (1+a_w^{-1})  /2\sqrt{i\omega t_2}$ 
for $\omega \ll  1/t_1' $ and $R_Q \cong 1/4i\omega\sqrt{t_1 t_2}$ 
for $\ 1/t_1' \ll \omega\ll c/L$.

In this situation we may calculate the heat flux $\dot{Q}_L$ 
at $x=L$. From Eq.(2.26) it is written  as 
\be 
\dot{Q}_L= \frac{(1-Z)e^{-ikL}}{1-Ze^{-2ikL}}\dot{Q}_0, 
\en 
which vanishes for $Z=1$ (or for $a_w=0$) 
and becomes small with increasing $\delta_BL$. 
Near the resonance frequency $n\pi c/L$,   
the ratio $\dot{Q}_L/\dot{Q}_0$  behaves as $(-1)^nW/
[i(\omega L/c-n\pi)+ n\pi\zeta/2\rho c^2+ W]$. 
The low frequency behavior for $\omega\ll c/L$ 
is given by 
\be 
\dot{Q}_L= (1+ 2\sqrt{i\omega t_1}X_v)^{-1} \dot{Q}_0.  
\en 
From the discussion  below Eq.(3.16),  we find 
$\dot{Q}_L \cong \dot{Q}_0$ 
for $\omega t_1'  \ll 1$ under Eq.(3.17).
That is,  an applied heat flux passes through 
a near-critical fluid on the time scale of $t_1'$ 
under Eq.(3.17),  due to the piston effect.

\subsection{Thermal and mechanical piston effects}

\begin{figure}[th]
\includegraphics[scale=0.37]{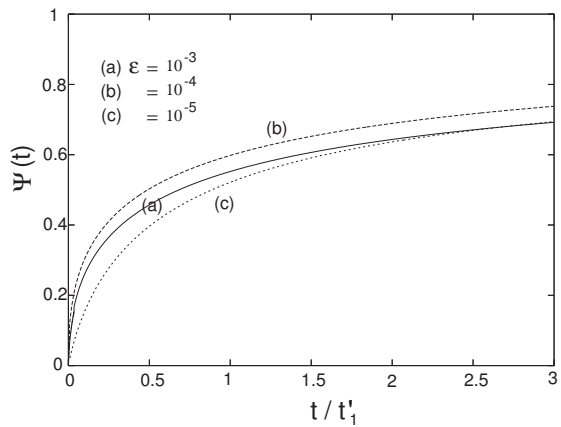}
\caption{Relaxation function $\Psi(t)$ in 
Eq.(3.29)  defined by Eq.(3.30) 
vs  $t/t_1'$ 
 at  $\epsilon=10^{-3},$ $10^{-4}$, and $10^{-5}$ 
   for CO$_2$ in a Cu cell with 
$L=1$ cm, which is applicable for $t\gg L/c$.  
The time $t_1'$ is defined in Eq.(3.16). Functional 
form of $\Psi(t)$ as a function of $t/t_1'$  
is rather insensitive to $\epsilon$ 
since  Eq.(3.17) holds for this case. 
 }
\end{figure}

\subsubsection{Boundary temperature change}

In the original 
papers of the piston effect \cite{Ferrell}, the 
boundary temperatures at $x=0$ and $L$ 
were both  raised by a common 
small amount  $T_1$ at $t=0$. 
Subsequently, the boundary temperatures were held fixed 
for $t>0$.  In this paper, we examine the effects  of 
 finite $a_w^{-1}$ \cite{Hao,Moldover} 
and large $\zeta$ \cite{Carles0,Carles}. 
We suppose  that 
 the system was in equilibrium for $t<0$ 
and the temperature in the wall region $x<0$ 
was instantaneously raised by $T_1$ at $t=0$ 
without external heat input  in the other wall region $x>L$.  
The  boundary conditions  are then given by 
$\delta T (x,t) \rightarrow T_1$ 
as $x\rightarrow -\infty$ 
and  $\delta T (x,t) \rightarrow 0$ 
as $x-L\rightarrow \infty$  for $t>0$. 
All the deviations vanish for $t<0$.

The Fourier transformation of the interior temperature deviation 
$\delta T(x,t)$ with respect to $t$  is given by Eq.(3.13) 
with $T_w=T_1e^{i\omega t}/i\omega$ (since $\int_0^\infty dt e^{-i\omega t}=1/i\omega$).  
The  inverse Fourier transformation gives 
\be 
\frac{\delta T(x,t)}{T_1} 
 =   \int \frac{d\omega  e^{i\omega t}}{2\pi i\omega}
( e^{-ikx}+  Ze^{ikx-2ikL})R_T, 
\en     
where the integration is in the range $[-\infty,\infty]$.
Under Eq.(2.34),  
$W=W(\omega)$ and $R_T=R_T(\omega)$ are given by Eqs.(2.40) 
and (3.12), respectively. 
The integrand is analytic (without singularities) 
in the lower half plane 
 ${\rm Im}\omega<0$ and hence 
the integral is nonvanishing only for $t>0$.

In the time region $t\gg L/c$ we may 
 neglect the space dependence of 
$\delta T(x,t)$ in the interior and use the simple expression (3.15) 
for $R_T(t)$. It then follows   
\be 
\delta T(t) =  T_1 \Psi(t)/2,
\en 
where we introduce the  dimensionless relaxation function $\Psi(t)$. 
Its  Fourier transformation  reads  
\be 
\int_0^\infty  dt e^{-i\omega t} \Psi(t)= 
\frac{1}{i\omega(\sqrt{i\omega t_1}X_v+1)}. 
\en 
The inverse Fourier 
transformation of the right hand side 
of Eq.(3.30) may   be transformed into an integral  
along the positive imaginary axis ${\rm Im}\omega>0$.   
With $X_v$ being defined by Eq.(2.41), we generally find 
$\Psi(t)>0$ for $t>0$,  
$\Psi(t) \cong t/\sqrt{t_1t_B}$ 
as $t\rightarrow 0$,  and 
$\Psi (t)= 1- (t_1'/\pi t)^{1/2}+\cdots$ as $t\rightarrow \infty$. 
In  particular, not very close to the critical point, 
we may neglect the bulk viscosity 
 and take   the limit $t_B\rightarrow 0$; then, 
  $X_v\rightarrow 1+a_w^{-1}$ and 
$\Psi (t) \rightarrow \Psi_0(s)$, where $\Psi_0(s)$ is 
a universal function of $s=t/t_1'$ expressed as \cite{Ferrell}  
\be 
\Psi_0(s)= 1- \int_0^\infty 
\frac{d u}{\pi\sqrt{u}} \frac{e^{- us}}{1+u}=1-e^{s}
{\rm erfc}(\sqrt{s}), 
\en  
where ${\rm erfc}=1-{\rm erc}$ is the complementary 
 error function  and $\Psi_0\cong 2(s/\pi)^{1/2}$ 
for $s\ll 1$ and  
$\Psi_0\cong 1- 2(\pi s)^{-1/2}$ 
for $s\gg  1$.

In Fig. 6, we display $\Psi(t)$ as a function of $t/t_1'$ 
at   $\epsilon=10^{-3},$ $10^{-4}$, and $10^{-5}$ 
   for CO$_2$ in a Cu cell with 
$L=1$ cm.   For  $\epsilon=10^{-3}$ we can see 
 $\Psi(t)\cong \Psi_0(t/t_1')$, where 
 $t_B/t_1' \sim 0.02$ from Table 1. 
The discussion below Eq.(3.16) 
indicates  that $\Psi(t)$ approaches  unity 
on the time scale of $t_1'$ as long as Eq.(3.17) is satisfied. 
This is the case even for  $\epsilon=10^{-5}$, where 
 $t_B/t_1'\cong  21$ from Table 1. In fact, if  $t_B/t_1'\gg 1$ 
and $a_w\ll 1$, we may set $\sqrt{i\omega t_1} 
X_v \cong i\omega \sqrt{t_1t_B} + a_w^{-1}\sqrt{i\omega t_1}$,  
where the second term is relevant  
  in  $\Psi(t)$ under  Eq.(3.17),  again leading to 
$\Psi(t) \cong \Psi_0(t/t_1')$ for $t>a_w^2t_B$.
However,  the reverse condition of Eq.(3.17) holds extremely 
close to the critical point, where 
$R_T \cong 1/[i\omega \sqrt{t_1t_B}+1]$ 
holds yielding  \cite{Carles} 
\be 
\Psi(t) \cong  1- \exp(-t /\sqrt{t_1t_B}).
\en 
The new relaxation time $\sqrt{t_1t_B}$ here grows 
as $\sqrt{t_1t_B}\cong 1.0\times 10^{-4}L\epsilon^{-0.64}$sec
  for near-critical CO$_2$.

Assuming  the isothermal boundary ($a_w=\infty$), 
Carl$\grave{\rm{e}}$s 
 and Dadzie examined the bulk viscosity effect 
in the thermal equilibration \cite{Carles}. 
Their relaxation function is 
obtained if we set $X_v=({1+i\omega t_B})^{1/2}$ in Eq.(3.30). 
Then  a  new  viscous regime appears 
for $t_1\gg t_B$  with 
 $\Psi(t)$ being given by Eq.(3.32), while 
the usual piston regime is encountered 
for $t_1\ll t_B$. For CO$_2$ we 
have $t_B/t_1\cong 2.7 \times 10^{-20}L^{-2}
\epsilon^{-4.63}$, so 
$t_1=t_B$ holds at $\epsilon\cong 0.6\times 10^{-4}$ with 
 $L=1$cm.   In our calculations based on Eq.(3.17),  
the different predictions have arisen 
from the reduced temperature dependence of 
$a_w$ or the crossover of the boundary 
condition into the insulating one.

 \subsubsection{Volume change}

We suppose a  volume change 
by moving  the boundary wall at $x=0$ 
 by a small length 
$u_1$ instantaneously  at $t=0$ \cite{Onukibook}. 
We assume the thermal boundary conditions (2.28) and (2.30) 
at $x=0$ and $L$. 
As in Eq.(3.26), the complete interior density deviation 
is the inverse Fourier transformation of Eq.(3.8), where $\beta$ is 
given by Eq.(3.20)  with  $u_w=u_1e^{i\omega t}/i\omega$.

Here we are interested in the late stage 
 $t\gg L/c$, where  the interior  deviations  depend only on  $t$. 
The inverse Fourier transformation of Eq.(3.21) gives  
the interior  deviations, 
 \be 
\delta \rho(t)= b_s \delta T(t) 
 = [1-\Psi(t)]\rho u_1/L , 
\en 
where $\Psi(t)$  defined by Eq.(3.30) represents the effect 
of the thermal diffusion layers at $x=0$ and $L$. 
The above form with $\Psi=\Psi_0$ 
was derived in Ref.\cite{Onukibook}.
If $u_1>0$, the interior is adiabatically heated by 
$b_s^{-1}\rho u_1/L$ on the acoustic time scale 
$L/c$ after the volume change,  
 while the boundary wall temperature is almost unchanged.  
Subsequently, the thermal diffusion layers 
become effective as {\it reverse}  pistons 
and  the interior temperature  deviation 
decays as  $(t_1'/t)^{1/2}$.

The reverse piston effect itself generally occurs 
on the time scale of $t_1'$ after 
a  near-critical fluid was adiabatically 
heated or cooled.  Miura {\it et al.} 
observed such a process after a pulse-like heat input   
(see Fig. 2 in Ref.\cite{Miura}).

\subsection{ Emission of sound }

We examine sound emission at the boundary at $x=0$.  
We  neglect  the 
incoming wave reflected  at the other end $x=L$ 
and consider the semi-infinite limit  
$L \rightarrow \infty$.

The problem is simple in 
 the case of boundary wall motion. An emitted  sound  propagates   
with the velocity   $c$  and  integration of 
 the continuity equation 
gives the density deviation, 
\be 
\delta\rho (x,t)\cong  \rho v_1(t-x/c)/c,
\en 
where $v_1(t)$ is the velocity of the boundary. 
The localized part of the density deviation 
(the term proportional to $a$ in Eq.(2.20)) 
should be small when differentiated with respect to time. 
In fact, under the adiabatic condition  
(2.34),  Eqs.(3.12) and (3.20) lead to  
\be 
\beta\cong \frac{1+Z}{2(1+\Delta_v)c}
i\omega\rho u_w,
\en  
for $e^{-2ikL}\rightarrow   0$. If we 
set ${1+\Delta_v}\cong 1$ and $Z\cong 1$,  the above  relation 
becomes $\beta \cong i\omega \rho u_w/c$,  
leading to Eq.(3.34).  Here  $i\omega u_w$ is  the Fourier 
transformation of $v_1(t)$ multiplied by $e^{i\omega t}$. 
Thus Eq.(3.34) holds 
on time scales longer than 
$t_\xi$ (even when the time scale of $v_1(t)$ 
is shorter than $t_B$).

A sound is also emitted  when   
  a time-dependent 
heat flux $\dot{Q}_0(t)$ 
is supplied at the boundary at $x=0$. 
From Eq.(3.8) the Fourier transformation of 
 the interior density deviation $\delta\rho (x,t)$ 
is of the form $\beta e^{ikx}$ with 
$\beta$ being given by Eq.(3.23). 
Under the adiabatic condition Eq.(2.34) 
we may use the second line of Eq.(3.24) to find the 
convolution relation, 
\be 
\delta \rho(x,t) 
=  \frac{\rho}{cT}\ppp{T}{p}{s} \int_{-\infty}^t  d\tau
\Phi(x,t-\tau){\dot Q}_0(\tau).
\en   
The memory function  $\Phi(x,t)$ is defined for $t>0$ as  
\be 
\Phi(x,t)= 
\int \frac{d\omega}{2\pi} 
e^{i\omega t-ikx}(1+\Delta_v)^{-3/2},  
\en    
where  $\Delta_v= i\omega \zeta/\rho c^2=i\omega R_B t_\xi$ 
with $R_B\cong 0.03$.  The time integration 
of this function is normalized as $\int_0^\infty dt 
\Phi(x,t)=1$. From the integration in the 
region $\omega <t_\xi^{-1}$ 
we obtain  the  long-time behavior  
$\Phi(0,t)\cong 
 (4t/{\pi} t_\zeta^{3})^{1/2}e^{-t/t_\zeta}$ 
with $t_\zeta\equiv R_Bt_\xi$ at $x=0$. 
Since this relaxation is rapid, 
we may set $\Phi(0,t) \cong \delta(t)$ 
($\delta$-function) at $x=0$ on time  
scales longer than $t_\xi$ or when ${\dot Q}_0(t)$ varies 
slower than $t_\xi$.    Furthermore, 
if the distance $x$ is not large such that 
the bulk damping is negligible in the region $0<x<L$, 
we may set $\Phi(x,t) \cong \delta(t-x/c)$ to find 
the simple formula for the emitted sound,   
\be 
\delta \rho(x,t) 
=  \frac{\rho}{cT}\ppp{T}{p}{s} {\dot Q}_0(t-x/c),
\en 
as the counterpart of Eq.(3.34). 
 On the other hand, use of Eq.(B6) for $\Delta_v$ gives  
the short-time behavior, 
\be 
\Phi(0,t)=
({\hat{\alpha}}/{2\hat{\nu}}) 
(t/t_\xi)^{{\hat{\alpha}}/{2\hat{\nu}}}/t, 
\en   
valid in the time region   $t\ls t_\xi$ with 
${\hat{\alpha}}/{2\hat{\nu}}\cong 0.088 $ (see Appendix B). 
This behavior is detectable  only for an increase of 
$\dot{Q}_0$ within a time shorter  than $t_\xi$.

Miura {\it et al.}  applied a stepwise 
heat flux with ${\dot Q}_0 =0.183\times 10^7$   
to find a stepwise outgoing sound  
with $\delta\rho/\rho \cong 
 2.2\times 10^{-7}$ for CO$_2$, 
 where  $\dot{Q}_0$ is in cgs units (erg$/$cm$^2$sec) \cite{Miura}. 
Our theoretical expression (3.38) 
becomes  $\delta \rho/\rho = 1.38\times 10^{-13}{\dot Q}_0$ 
with the aid of 
$(\p T/\p p)_s\cong T_c/6.98p_c$ for CO$_2$ 
\cite{Hohenberg}. 
For their  experimental $\dot{Q}_0$  our theory  gives  
$\delta \rho/\rho = 2.55\times 10^{-7}$ 
in fair agreement  with  
the observed density change. Furthermore, they could generate 
sound pulses with duration of order $10 \mu$sec by 
applying  short-time  heat input. 
They were interested in the adiabatically 
increased energy $E_{\rm ad}\equiv 
{p}\int dx \delta\rho(x,t)/\rho$ in 
 the pulse region per unit area. Here 
Eq.(3.38) yields \cite{Ferrell} 
\be 
E_{\rm ad}= \frac{p}{T}\ppp{T}{p}{s} 
{ Q},
\en 
where $Q= \int dt {\dot Q}_0(t)$ is 
 the total heat 
supplied. The ratio $E_{\rm ad}/{Q}$  
 represents 
the efficiency of transforming applied  heat to mechanical 
work.  Theoretically, it is  given by  $(\p T/\p p)_sp/T$ 
as in Eq.(3.40) and is equal to 
$1/6.98= 0.14$  for near-critical CO$_2$ \cite{Hohenberg}.  
The measured values of the ratio 
$E_{\rm ad}/{Q}$  
were in the range  $0.11-0.12$ again in fair 
agreement with our theory.

\subsection{Reflection of sound }

Reflection of plane wave sounds 
is discussed for an isothermal boundary 
in the textbook of 
Landau-Lifshitz \cite{Landau}. 
Miura {\it et al.} \cite{Miura} observed reflected 
pulses  passing  through a detector   
in the cell. Their shapes 
gradually flattened after many traversals 
within  the cell, resulting in  the interior 
temperature homogenization. 
At present, it is not clear how  to understand their data. 
Here, as a first step, 
 we will derive  some fundamental relations 
on sound  reflection.

We  consider a pulse approaching to the boundary at $x=0$ 
 in the semi-infinite limit $L\rightarrow \infty$. 
Reflection takes place upon its encounter with the wall. 
The density deviations of the 
incoming and outgoing pulses are obtained as 
the inverse Fourier transformation of Eq.(2.20). 
Neglecting  the bulk damping in the neighborhood of the boundary,  
we may  express them  as 
$\rho_{i}(t+x/c)$ and $\rho_{o}(t-x/c)$, respectively. 
Using   $\alpha(\omega)= e^{i\omega t}\int d\tau e^{-i\omega \tau} 
\rho_{i}(\tau)$ and   $\beta=Z\alpha$, we obtain 
\be  
\rho_{o}(t)=
 \int \frac{d\omega }{2\pi}\int dt' 
Z(\omega)
e^{i\omega (t-t')}\rho_{i}(t').   
\en
The interior  density deviation is the sum $\delta\rho(x,t)= 
\rho_{i}(t+x/c)+\rho_{o}(t-x/c)$. 
Since $Z(0)=1$, 
the excess mass is invariant upon reflection as 
\be 
\Delta M= \int dt \rho_{i}(t) =\int dt \rho_{o}(t). 
\en 
This  relation holds 
if we integrate a long tail of 
 the reflected pulse $\rho_o(t) $  
at large $t$ (see Eq.(3.47)).

If $\rho_{i}(t)$ changes much slower than $t_\xi$, 
 we may set $Z\cong 1-2W$ 
with $W= (\gamma-1)\sqrt{i\Delta_T}/X_v$ 
from  Eq.(2.40).  In this approximation we may rewrite 
Eq.(3.40) in the following convolution form, 
\bea 
\rho_{o}(t) &=&\rho_{i}(t)- 
 \int_0^\infty d\tau \dot{\chi} (\tau) 
[\rho_{i}(t-\tau) -\rho_{i}(t)]\nonumber\\
&=&\rho_{i}(t)- 
 \int_0^\infty d\tau {\chi} (\tau) 
\dot{\rho}_{i}(t-\tau)   
\ena
where $\dot{\chi}(t)= \partial \chi(t)/\partial t$ 
and $\dot{\rho}_{i}(t)= \partial \rho_{i}(t)/\partial t$. 
From Eq.(3.41) the function $\chi(t)$ is the inverse Fourier 
transformation of $(1-Z)/i\omega \cong 2W/i\omega$. 
Some calculations (in the complex $\omega$ plane) 
give $\chi(t)$ in  
the integral form,
\be
{\chi(t)}={\varepsilon_r}  
\int_0^\infty \frac{d\Omega}{\pi\sqrt{\Omega}}
  {\rm Re}\bigg[ \frac{e^{-\Omega s}
}{a_w^{-1}+\sqrt{1-\Omega}}\bigg],
\en
where $s=t/t_B$ is the scaled time, ${\rm Re}[\cdots]$ 
denotes taking the real part,  and $\sqrt{1-\Omega}=i\sqrt{
\Omega-1}$ for $\Omega>1$. The dimensionless parameter 
$\varepsilon_r$ is defined by 
\be 
\varepsilon_r= 2(\gamma-1)\sqrt{D/c^2t_B}
\en 
which decreases  near the critical point as 
$\varepsilon_r \cong 4.3\epsilon^{0.75}$ for CO$_2$.  
 The function  $\chi(t)$ depends only on $s$ and $a_w$. 
For $a_w\gg 1$ we have ${\chi(t)}/{\varepsilon_r}
\cong e^{-s/2}I_0(s/2)$ with $I_0$  being the modified 
Bessel function, while for $a_w\ll 1$ 
we have ${\chi(t)}/{\varepsilon_r}
\cong 1-\Phi_0(s/a_w^2)$ 
 with $\Phi_0$  being defined in Eq.(3.31). 
Thus $\chi(t)$ changes 
 on the scale of $t_B'\equiv  
t_B (1+a_w^{-1})^{-2}$ and its limiting 
behaviors  are as follows: 
\bea 
\frac{\chi(t)}{\varepsilon_r} &=&
(1+a_w^{-1})^{-1} (\pi s)^{-1/2}+\cdots 
\quad (s \rightarrow \infty) ,\nonumber\\
&=&1-2a_w^{-1}(s/\pi)^{1/2}+\cdots \quad 
(s\rightarrow 0) .
\ena 
In addition, the  second term of Eq.(3.43) 
representing the distortion 
is negative (positive) when $\rho_i(t)$  
is increasing (decreasing). 
This  initial drop is because of 
 heating and expansion of the  pulse 
at the boundary.

From the first line 
we  obtain $\dot{\chi}(t) \cong -(t_2/\pi)^{1/2}t^{-3/2}$ 
for $t\gg t_B'$. (i) Let $\rho_i(t)$ is peaked in the region 
$|t|<t_0$; then, for $t\gg t_B'$ and $t_0$, 
the first line of Eq.(3.43) gives a long-time tail 
of the reflected wave, 
\be 
[\rho(t)]_{\rm tail}= \Delta M 
(t_2/\pi)^{1/2}t^{-3/2}
\en 
where $\Delta M$ is defined by Eq.(3.42). 
If $t_0>t_B'$,   the total mass  
behind the peak  is  given by 
 the time integral of the tail Eq.(3.47) 
in the region $[t_0,\infty]$. Thus the mass  fraction 
behind the peak is $(4t_2/\pi t_0)^{1/2}$.   
For CO$_2$ this quantity is estimated as 
$10^{-7}\epsilon^{-0.75}t_0^{-1/2}$ 
with $t_0$ in sec for $a_w \gg 1$. 
(ii) As another example, we consider a stepwise 
change, where  $\rho_i(t)$ 
is   equal to 
0 for $t<0$ and to a constant $\rho_1$ 
for $t>t_0$ with $t_0$ being the transient   time. 
 Then, for $t\gg t_B'$ and $t_0$, 
the second  line of Eq.(3.43) gives a longer  tail,
\be 
[\rho(t)]_{\rm tail}= \rho_1  
(4t_2/\pi)^{1/2}t^{-1/2}. 
\en 
The bulk viscosity does not appear in these tails.

When $\rho_{i}(t)$ changes much slower than $t_B'$, 
only the long time behavior of $\chi(t)$ is relevant in $\rho_o(t)$. 
From Eq.(3.43) we find the following convolution relations,   
\bea 
\rho_{o}(t)&=&\rho_{i}(t) 
+ \sqrt{\frac{t_2}{\pi}}\int_0^\infty \frac{d\tau}{\tau^{3/2} }
[\rho_{i}(t-\tau) -\rho_{i}(t)] \nonumber\\ 
&=&\rho_{i}(t) 
- \sqrt{\frac{4t_2}{\pi}}\int_0^\infty \frac{d\tau}{\sqrt{\tau} }
\dot{\rho}_{i}(t-\tau) , 
\ena 
from which the long-time tails (3.47) and (3.48) 
readily follow. 
The above expressions contain only $t_2$ in Eq.(2.43) 
and  not $t_B$. They are widely  applicable  far from the critical 
point (where $t_B$ becomes short).  
With decreasing  $\epsilon$ for the 
isothermal boundary ($a_w>1$), 
 $t_2$ grows  and 
the distortion of the reflected pulse increases 
 as long as the pulse width is longer 
than $t_B$.  However, if the pulse width is shorter than 
$t_B'$,  the distortion decreases on approaching the 
critical point since 
$\varepsilon_r$  in Eq.(3.45) decreases.

\begin{figure}[t]
\includegraphics[scale=0.4]{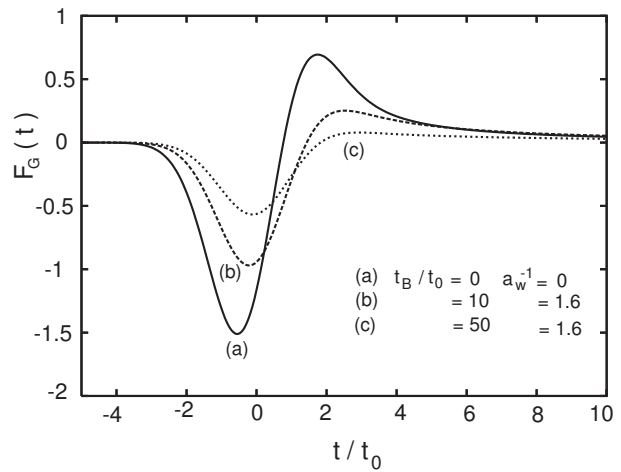}
\caption{Scaled pulse  deformation 
$F_G(t)$ defined by Eq.(3.50) vs 
scaled time $t/t_0$ 
for a Gaussian  incoming pulse 
with width $t_0$ for 
$t_B/t_0\rightarrow 0$ 
and $a_w=\infty$  (a), 
 $t_B/t_0= 10$ and $a_w=0.63$  (b), and  
$t_B/t_0= 50$ and $a_w=0.63$  (c). 
}
\end{figure} 

As a simple  illustration, let us 
consider a Gaussian pulse 
$\rho_{i}(t)=\rho_1\exp(-t^2/2t_0^2)$,  
where $\rho_1$ is the pulse height and 
$t_0$ is  the pulse width. 
Since its  Fourier transformation is $(2\pi)^{1/2}
\rho_1t_0e^{-\omega^2t_0^2/2}$, we may readily calculate 
$\rho_{o}(t)$. In Fig. 7, we plot the normalized 
pulse deformation defined by  
\be
F_G(t)= [{\rho_{o}(t)}-\rho_{i}(t)]/(\rho_1\sqrt{t_2/t_0}) .
\en 
The curve (a) is for the limiting case 
$t_B/t_0\rightarrow 0$ 
and $a_w=\infty$, while 
 $t_B/t_0= 10$ and $a_w=0.63$  in (b), and  
$t_B/t_0= 50$ and $a_w=0.63$  in (c). 
In Table 1 we have  $a_w=0.63$ and 
$t_B= 1.9$msec at $\epsilon=10^{-4}$ for 
CO$_2$ in a Cu cell, where  pulses with $t_0 \ll 
t_B$ are well possible \cite{Miura}. 
We recognize that the distortion is negative 
for $t\ls t_0$ and is positive for $t\gs t_0$ 
(in accord with the comment below Eq.(3.43)) 
and that the distortion  is decreased 
as $t_B/t_0$ is increased 
or for shorter pulses due to the bulk viscosity growth.

\section{Summary and remarks}

In summary, we have examined various thermoacoustic effects 
in one-component supercritical fluids in a one-dimensional 
geometry.  We summarize our main results.\\
(i) In the linear hydrodynamics,  sound modes 
and thermal diffusion modes 
are both present as in Eq.(2.20),  
depending on given boundary conditions. 
The latter modes can be absent only for the insulating 
boundary condition $a_w=0$. 
The calculations are straightforward and 
the final expressions are much simplified 
under the adiabatic condition (2.34) 
or for low frequencies $\omega\ll c^2/\gamma D$.  
It is remarkable that the bulk viscosity $\zeta$ 
appears in the combination $\omega \gamma \zeta/\rho c^2=
\omega t_B$ as first pointed out by Carl$\grave{\rm e}$s 
\cite{Carles0,Carles}. 
The resultant characteristic time 
$t_B$ grows as $\epsilon^{-3.0}$, while 
the life time of the critical fluctuations $t_\xi$ grows as 
$\epsilon^{-1.9}$. 
\\
(ii) We have  introduced the reflection 
factor $Z(\omega)$ as the ratio between 
 outgoing and incoming sounds. 
Using  $Z$ or $W=(1-Z)/(1+Z)$ we 
have examined  the acoustic eigenmodes, 
the response of the fluid to 
 applied oscillation of the boundary 
temperature, the boundary  heat flux, and the boundary position. 
To these thermal and mechanical 
perturbations, resonance is induced 
 when the  frequency of the perturbation 
is close to one of the eigenfrequencies, while 
nearly uniform adiabatic changes are caused in the interior 
at much lower  frequencies  owing to the piston effect. \\
(iii) 
We have also examned the response to a stepwise 
change of the boundary temperature and the boundary position. 
The relaxation time is given by the modified piston time 
$t_1'$ in Eq.(3.16) first 
introduced by Ferrell and Hao \cite{Hao}. 
It is  equal to the original piston time $t_1$ 
in Eq.(1.1) for the isothermal boundary 
 $a_w\gg 1$ and to $a_w^{-2} t_1$ for the 
insulating boundary $a_w \ll 1$.\\ 
(iv) 
As the critical point is approached,  
 the role of the thermal diffusion layers 
is eventually diminished both by  
 decreasing of the effusivity ratio $a_w$ 
and by  growing of the bulk viscosity $\zeta$, 
while the bulk sound attenuation 
becomes increasing stronger.  The bulk viscosity effect 
in the thermal diffusion layer 
is thus masked by its  enhanced effect in the bulk. 
\\
(v) For CO$_2$ in a Cu cell, 
the  boundary becomes thermally  insulating 
for $\epsilon \ll 10^{-4}$.    This  suppresses   
the bulk viscosity effect in the 
thermal diffusion layers as 
long as Eq.(3.17) holds 
or for $\epsilon>10^{-6}$. 
In this case, the viscous regime predicted by 
 Carl$\grave{\rm e}$s and Dadzie emerges  for $\epsilon<10^{-6}$ 
\cite{Carles0,Carles}.  To increase this crossover reduced 
temperature, the cell length $L$ needs to be shorter. 
 For the wall materials in 
Ref.\cite{Moldover}, this crossover occurs much closer 
to the critical point \cite{effusive}.  
\\ 
(vi) We have also examined  
sound emission and   reflection  at the boundary, 
which are elementary hydrodynamic processes but 
seem to have not been  well examined \cite{Landau}. 
For emission,  the   formulas (3.34) and (3.38) 
are valid for a mechanical piston and a thermal 
heat input on  
time scales longer than $t_\xi$.
For reflection,  Eq.(3.43)  with Eq.(3.44) 
holds on  time scales longer than $t_\xi$. 
The formula (3.49)  is the classical 
one  valid on time scales 
much longer than $t_B$, where 
the distortion of the outgoing 
pulse  increases on approaching 
the critical point.   For pulses shorter than $t_B$, 
the distortion of the outgoing pulse is decreased 
as can be seen in Fig. 7.
\\

In this paper, we have treated 
near-critical  fluids   in one phase states. 
However, more challenging 
are hydrodynamic effects in 
in two phase states, where latent heat transport, wetting dynamics,  
and Marangoni convection 
come into play in addition to the piston effect 
\cite{Beysens1,Beysens2,OnukiV}.

\begin{acknowledgments}  
I would like to thank 
the members of the experimental group 
of the piston effect in Japan \cite{Miura},  
Takeo Satoh, and P. Carl$\grave{\rm{e}}$s  
for valuable discussions.  
Thanks are also due to Horst  Meyer, 
K.A. Gillis,  and M.R. Moldover 
for informative correspondence. 
This work was  
 supported by   grants from the Japan Space Forum 
 and from 
the 21st Century COE project (Center for Diversity and Universality in
Physics) from the Ministry of Education, Culture, Sports, Science and 
Technology of Japan.

\end{acknowledgments}

\vspace{2mm} 
{\bf Appendix A: Simple theory of  the piston effect}\\
\setcounter{equation}{0}
\renewcommand{\theequation}{A\arabic{equation}}

Here we  give a simple derivation of $t_1$ in Eq.(1.1). 
Let us apply a small heat $\delta Q$ 
to a fluid from the boundary per unit area 
in the one dimensional geometry. The volume expansion 
of the thermal diffusion layers 
is given by 
$(\p T/\p p)_s  A{{\delta Q}(t)}/{ T},
$ 
where $A$ is the area of the heater surface and 
use is made of the Maxwell relation 
$({\p \rho^{-1}}/{\p s})_{p} =(\p T/\p p)_s$. 
The interior density change and the 
pressure change $\delta p$  are nearly 
homogeneous in the interior and are given by 
\be 
 \delta\rho= \frac{\delta p}{c^2}= \frac{\rho}{T}
\ppp{T}{p}{s}\frac{\delta Q}{L}.
\en 
where $L=V/A$ is the cell length. 
The interior temperature deviation is caused adiabatically 
as 
$\delta T= (\p T/\p p)_s\delta p$ and is written as  
\be 
\delta T =  (\gamma-1) \delta Q/ C_p L,
\en
where $C_p= \rho T (\p s/\p T)_p$ 
is the isobaric specific heat and use is made of Eq.(2.9). 
If the boundary temperature is raised 
by $T_1$ at $t=0$,   we have 
  $\delta Q\sim   C_p \ell(t)T_1$ in the early stage, 
where $\ell(t)= \sqrt{Dt}$ is the 
thickness of the thermal diffusion layer.
If we set $\delta T =T_1/2$, Eq.(1.1)  is reproduced.

The relation (A1) also follows from 
our formula Eq.(3.38). 
Let the heat input rate $\dot{Q}(t)$ from the boundary to the fluid  
change slowly compared to 
the acoustic time $t_a=L/c$. 
We suppose  a time interval with width 
$\delta t \gg t_a$, in which  $\dot{Q}(t)$ is almost 
unchanged.  Since $\delta t/t_a$ is the traversal number 
much larger than unity,   
the adiabatic pressure and  density 
increases in the interior region are given by 
\be 
 \delta\rho=
\frac{\delta p}{c^2}= 
\frac{\delta t}{t_a} 
\frac{\rho}{cT}\ppp{T}{p}{s} \dot{Q}, 
\en 
as a result of superposition of many steps. 
In terms of 
the incremental 
heat supply $\delta Q= \dot{Q}\delta t$
  we reproduce Eq.(A1).

\vspace{2mm} 
\vspace{2mm} 
{\bf Appendix B: Summary of critical behavior}\\
\setcounter{equation}{0}
\renewcommand{\theequation}{B\arabic{equation}}

Let a one-component fluid  be on the critical isochore  
($\rho=\rho_c)$ 
with small  positive  $\epsilon=  T/T_c-1$  
near the gas-liquid critical point. 
The physical parameters used 
in the table 1 and the figures are given below. 
  Hereafter 
$\hat{\nu}(\cong 0.63)$, $\hat{\gamma}(\cong 1.24)$, 
and $\hat{\alpha}(\cong 0.10)$ are the usual critical 
exponents.  Data of near-critical CO$_2$ 
can be found in Refs.\cite{Swinney,Hohenberg}.

 Our hydrodynamic description is valid when 
the spatial scale under investigation is longer than 
the correlation length  $\xi =\xi_0 
 \epsilon^{-\hat{\nu}}$, where $\xi_0=1.5 {\rm \AA}$ 
for CO$_2$.  
The constant-volume specific heat 
$C_V=\rho T(\p s/\p T)_\rho$ 
and the isobaric specific heat 
$C_p=\rho T(\p s/\p T)_p$ 
are  expressed as 
\be 
C_V = A_V[\epsilon^{-\hat{\alpha}}+B], \quad 
C_p = A_p\epsilon^{-\hat{\gamma}}.  
\en 
For CO$_2$ on the critical isochore,  the coefficients 
are given by 
$A_V= 26.3 k_B n^*$, $B=0.9$, and 
$A_p= 2.58k_B n^*$, where 
 $n^*= p_c/k_BT_c\cong 1.76\times 10^{21}$cm$^{-3}$. 
The specific-heat 
ratio $\gamma$ grows strongly as 
$\gamma_0 \epsilon^{-\hat{\gamma}+\hat{\alpha}}$ 
if the background ($\propto B$) is  neglected, 
where $\gamma_0=0.1$ for CO$_2$. 
 The sound velocity 
and the constant-volume  specific heat are 
weakly singular as $c^2 \propto \epsilon^{\hat{\alpha}}$ 
(if the background is neglected \cite{Onukibook}).  
We have set  $c=2.3\times 10^4\epsilon^{0.06}$ 
cm sec$^{-1}$ for CO$_2$ \cite{Miura}.

The thermal conductivity $\lambda$ grows 
such that the thermal diffusion constant $D$ 
behaves  as 
\be 
D=\lambda/C_p= 
k_BT/6\pi\eta\xi=D_0\epsilon^{\hat{\nu}}, 
\en 
where $D_0=4.0\times 10^{-4}$cm$^2$sec$^{-1}$ for CO$_2$. 
Thus $\lambda\propto \epsilon^{\hat{\nu}-\hat{\gamma}}$. 
The relaxation time of the critical fluctuations with size $\xi$ 
increases as 
\be 
t_\xi= \xi^2/D= t_0 \epsilon^{-3\hat{\nu}},  
\en 
where $t_0=0.56\times 10^{-12}$sec for CO$_2$. 
The shear viscosity $\eta$ 
is only  weakly singular and may be treated as a constant 
independent of $\epsilon$ and $\omega$ 
to make rough estimates. However, the 
 zero-frequency bulk viscosity $\zeta$  
  grows very strongly as 
\be 
\zeta=\rho c^2 R_B t_\xi, 
\en
where  $R_B$ is a 
universal number estimated to be about  
$0.03$ \cite{Onukibook,Onuki97}.    For CO$_2$, 
 $\zeta/\rho \cong 0.9\times 10^{-5}
\epsilon^{-2+2\hat{\alpha}}$cm$^2$sec$^{-1}$, 
so $\zeta/\rho D= \Delta_v/\Delta_T 
 \cong 0.02\epsilon^{-2-\hat{\nu} 
+2\hat{\alpha}}$ (see Eqs.(3.13) and (3.14)). 
  In the high frequency regime 
$\omega t_\xi \gg 1$, the complex sound velocity in Eq.(2.39) 
becomes  asymptotically independent of $\epsilon$ 
\cite{FB1,Onuki97,FolkMoser}.   Thus,   
 \be 
c^*(\omega) \cong c (i\omega  t_\xi)^{\hat{\alpha}/6\hat{\nu}}.  
\en 
Since the exponent ${\hat{\alpha}/6\hat{\nu}}$ 
is small,  we may set 
$\Delta_v= i\omega\zeta/\rho c^2\cong 
({\hat{\alpha}/3\hat{\nu}}) \ln(i\omega  t_\xi)$. 
Thus, in this  high frequency regime,  
${\rm Im} \Delta_v$   tends to the following universal number,  
\be 
{\rm Im} \Delta_v= \pi \hat{\alpha}/6\hat{\nu}\cong 0.27\times 2/\pi
\en 
In the high frequency regime $\omega t_\xi>1$, 
  $\Delta_v$ remains to be 
as a small quantity and 
the frequency-dependent bulk viscosity 
defined by $\zeta(\omega) \equiv \rho c^2\Delta_v/i\omega$ 
decays roughly  as $1/i\omega$ with increasing $\omega$.

Furthermore, in our thermoacoustic problems, 
we have introduced the time $t_B$ in Eq.(2.37), 
which behaves as  
\be 
t_B=t_B^0
\epsilon^{-3\hat{\nu}-\hat{\gamma}+\hat{\alpha}},
\en  
where $t_B^0= 1.7\times 10^{-15}$sec 
for CO$_2$.  The effusivity ratio 
$a_w$ in  Eq.(2.29) decreases  as 
\be 
a_w = a_w^0 
\epsilon^{\hat{\gamma}-\hat{\nu}/2}.
\en 
For $a_w^0\gg 1$ the boundary wall crosses over 
from  an isothermal 
one  to an thermally insulating one  
on approaching the critical point. 
For example, 
between Cu and CO$_2$, we have  
 $a_w^0=3\times 10^3$ \cite{Miura}, where  
 $a_w <1$ is reached for $\epsilon < 1.6\times 10^{-4}$.  
The $a_w^0$ was smaller for the walls used in 
Ref.\cite{Moldover,effusive}.



\begin{references}

\bibitem{Onukibook} A. Onuki, {\it Phase Transition Dynamics} 
(Cambridge University Press, Cambridge, 2002). 


\bibitem{Straub} 
K. Nitsche and J. Straub, Proc. 6th European Symp. on Material Science under
 Microgravity Conditions (Bordeaux, France, 2-5 December 1986); 
J. Straub and L. Eicher, Phys. Rev. Lett. {\bf 75}, 1554 (1995).


\bibitem{Ferrell} 
A. Onuki, H. Hao, and R. A. Ferrell, Phys. Rev. A {\bf 41}, 
2256 (1990); A. Onuki and R.A. Ferrell,
Physica A {\bf 164}, 245 (1990). 

\bibitem{Gammon} H. Boukari, J.N. Shaumeyer, M.E. Briggs,  and R.W. Gammon, 
Phys. Rev. A   {\bf 41}, 2260 (1990) ;
Phys. Rev. Lett. {\bf 65}, 2654(1990). 

\bibitem{Wilkinson} R.A. Wilkinson, G.A. Zimmerli, 
 H. Hao, M.R. Moldover, R.F. Berg, W.L. Johnson, 
R.A. Ferrell and  R.W. Gammon,  Phys. Rev. E   {\bf 57}, 436 (1998). 
\bibitem{Beysens} 
B. Zappoli, D. Bailly, Y. Garrabos,  B. Le Neindre, P. Guenoun 
and D. Beysens, Phys. Rev. A {\bf 41}, 2264 (1990);
P. Guenoun, B. Khalil, D. Beysens, Y. Garrabos, F.  Kammoun, B. Le Neindre, and B. Zappoli, Phys. Rev. E {\bf 47}, 1531 (1993);
Y. Garrabos, M. Bonetti, D. Beysens, F. Perrot, 
 T. Fr$\ddot{\rm o}$hlich, P. Carl$\grave{\rm e}$s   and B. Zappoli,    
 Phys. Rev. E {\bf 57}, 5665 (1998).

\bibitem{Be} R.P. Behringer, A. Onuki 
and H. Meyer, J. Low Temp. Phys. {\bf 81}, 71 (1990). 


\bibitem{Klein} H. Klein, G. Schmitz and D. Woermann, 
 Phys. Rev. A {\bf 43}, 4562 (1991). 

\bibitem{Straubp} 
J. Straub, L. Eicher and A. Haupt, Phys. Rev. E {\bf 51}, 5556 (1995);
J. Straub and L. Eicher, Phys. Rev. Lett. {\bf 75}, 1554 (1995).

\bibitem{Zhong} F. Zhong and H. Meyer
Phys. Rev. E {\bf 51}, 3223 (1995); 
A. Kogan and H. Meyer, J. Low Temp. Phys. {\bf 112}, 419 (1998).
\bibitem{Kogan} 
 F. Zhong, A. Kogan and H. Meyer, J. Low Temp. Phys. {\bf 108}, 
161 (1997). 


\bibitem{ZA} B.  Zappoli and A.D. Daubin, 
Phys. Fluids, {\bf 6}, 1929 (1995); 
D.  Bailly and  B.  Zappoli, Phys. Rev. E {\bf 62}, 2353 (2000). 
\bibitem{Maekawa} 
T. Maekawa, K. Ishii, M. Ohnishi and S. Yoshihara, 
Adv. Space Res. {\bf 29},  589 (2002); 
J.  Phys. A, {\bf 37}, 7955 (2004). 


\bibitem{Hao} R.A. Ferrell  and H. Hao,
Physica A {\bf 197}, 23 (1993). 



\bibitem{effusive} 
The effusivity is defined by 
$\epsilon_f= 
(C \lambda)^{1/2}=CD^{1/2}$ 
for each material, where $C$ is the isobaric specific heat 
per unit volume,  $\lambda$ is the thermal conductivity, 
and $D=\lambda/C$ is the thermal diffusivity.  
For Cu used in \cite{Miura},   
$\epsilon_f/k_B= 2.6\times 10^{23}$cm$^{-2}$sec$^{-1/2}$. 
In Ref.\cite{Moldover}, 
$\epsilon_f/k_B=4.6\times 10^{22}$ for 
a stainless steel resonator 
and $\epsilon_f/k_B=2.7\times 10^{21}$ for 
a polymer-coated resonator  in the same units. 
The diffusivity ratio is defined as in Eq.(2.29) in this paper. 


\bibitem{Carles0} 
P. Carl$\grave{\rm{e}}$s, 
Phys. Fluids, {\bf 10}, 2164 (1998).
   

\bibitem{Carles} P. Carl$\grave{\rm{e}}$s  and K. Dadzie, 
Phys. Rev. E {\bf 71}, 066310 (2005). 


\bibitem{Moldover} 
K.A. Gillis, I.I. Shinder, and M.R. Moldover, 
Phys. Rev. E {\bf 70}, 021201 (2004); {\bf 72},  
051201 (2005); 
K. A. Gillis, I. I. Shinder, and M. R. Moldover,
 Phys. Rev. Lett. {\bf 97}, 104502 (2006). 
In these papers they defined 
$\Delta_v=\omega\nu_{\ell}/c^2$ and $\Delta_T=\omega D/c^2$ 
without $i$. 



\bibitem{Ohnishi} 
M. Ohnishi, S. Yoshihara, M. Sakurai, Y. Miura, M. Ishikawa,
H. Kobayashi, T. Takenouchi, J. Kawai, 
K. Honda, and M. Matsumoto, 
Microgravity. Sci. Tech. XVI-1,306 (2005).


\bibitem{Miura} 
Y. Miura, S. Yoshihara, M. Ohnishi, 
K. Honda, M. Matsumoto, J. Kawai, 
M. Ishikawa, H. Kobayashi, and A. Onuki,  
Phys. Rev. E {\bf 74}, 010101 (R) (2006). 


\bibitem{Carlesnew} 
P. Carl$\grave{\rm{e}}$s, 
Phys. Fluids {\bf 18}, 126102 (2006).

 
\bibitem{Meyer} A.B. Kogan, D. Murphy and H. Meyer, 
 Phys. Rev. Lett.  {\bf 82}, 4635 (1999);  
A.B. Kogan and H. Meyer, 
Phys. Rev. E {\bf  63}, 056310 (2001). 



\bibitem{Azuma} 
H. Azuma, S. Yoshihara, M. Onishi, K. 
Ishii, S.  Masuda,  and T.  Maekawa, 
Int. J. of  Heat and  Mass  Transfer {\bf 42}, 
771 (1999).  


\bibitem{jet} 
T. Fr$\ddot{\rm{o}}$hlich, D. Beysens, and Y. Garrabos
Phys. Rev. E {\bf 74}, 046307 (2006).




\bibitem{Sakir0}  
 S. Amiroudine, , P. Bontoux, 
P. Larroud, B. Gilly  and B. Zappoli, 
J. Fluid  Mech.  {\bf 442}, 119 (2001).  




\bibitem{Chiwata} Y. Chiwata and A. Onuki, Phys. Rev. Lett. {\bf 87},
144301 (2001); 
A. Furukawa and A. Onuki, Phys. Rev. E  {\bf 66},
016302 (2002). 
 
\bibitem{Sakir} 
S.Amiroudine and B.  Zappoli, 
Phys. Rev. Lett. 90, 105303 (2003); 
G. Accary, I.  Raspo, P. Bontoux, and B. Zappoli, 
Phys. Rev. E {\bf 72}, 035301(R) (2005). 
 
\bibitem{Soboleva} 
E.B. Soboleva, 
Phys. Rev. E {\bf 68}, 042201 (2003). 

\bibitem{Accary} 
G. Accary, I. Raspo, P. Bontoux, and B. Zappoli, C.R. Mecanique, 
{\bf 332}, 209 (2004). 


\bibitem{Swinney}  H.L. Swinney  and D.L. Henry, 
Phys. Rev. A {\bf 6}, 2586 (1973). 




\bibitem{FB1} R.A. Ferrell and J.K. Bhattacharjee,
Phys. Lett. A {\bf 86}, 109 (1981); 
Phys. Rev. A {\bf 24}, R1643 (1981);
Phys. Rev. A {\bf 31}, 1788 (1985).



\bibitem{Onuki97} A. Onuki, {Phys. Rev. E}  {\bf 55},  403 (1997).


\bibitem{FolkMoser} 
 R. Folk and G. Moser,  Phys. Rev. E {\bf 57}, 683 
(1998); ibid.  {\bf 57}, 705 (1998).

\bibitem{Landau}  L.D. Landau and E.M. 
Lifshitz,  {\it Fluid Mechanics} (Pergamon, 1959). 

\bibitem{Hohenberg}  P.C. Hohenberg  and M. Barmartz, 
Phys. Rev. A {\bf 6}, 289 (1972). 





\bibitem{Beysens1} 
J. Hegseth, A. Oprisan, Y. Garrabos, 
V. S. Nikolayev, C. Lecoutre-Chabot, and D. Beysens 
Phys. Rev. E {\bf 72}, 031602 (2005). 

\bibitem{Beysens2} 
R. Wunenburger, Y. Garrabos, C. Lecoutre, D. Beysens,  J. Hegseth, 
F. Zhong, and M. Barmatz,  Int. J of Thermophysics {\bf 23}, 
103 (2002).


\bibitem{OnukiV} A. Onuki, {Phys. Rev. E} 
 {\bf 75},  036304 (2007).






\end{references}
\end{document}